\documentclass{ICSnewsletter}

\usepackage{graphicx}
\usepackage{xcolor}
\usepackage{amsmath}
\usepackage{tudacolors}
\usepackage{caption}
\usepackage{subcaption}
\usepackage{tuda-pgfplots}
\usepackage{xurl}
\usepackage{footmisc}
\usepackage{hyperref}
\usepackage{pgfplots}
\usepackage{lipsum}
\usepackage[table]{xcolor}
\usepackage[style=ieee]{biblatex}
\AtEveryBibitem{%
	\clearfield{related}
	\clearfield{month}
	\clearfield{day}
	\clearfield{pages}
    \ifentrytype{article}{%
        \clearfield{eprint}%
        \clearfield{eprinttype}%
        \clearfield{eprintclass}%
    }{}%
}
\DeclareSourcemap{
	\maps[datatype=bibtex]{
		\map[overwrite]{
			\step[fieldsource=eprinttype, match={arxiv}, final]
			\step[fieldset=institution, null]
			\step[fieldset=type, null]
		}
		\map[overwrite]{
			\pertype{book}
			\pertype{collection}
			\pertype{inproceedings}
			\pertype{incollection}
			\pertype{proceedings}
			\step[fieldset=series, null]
			\step[fieldset=volume, null]
		}
		\map[overwrite]{
			\step[fieldsource=shortjournal,fieldtarget=journaltitle]
		}
	}
}
\DeclareFieldFormat*{title}{#1}
\newcommand{\project}[1]{Project #1}
\newcommand{\nproject}[1]{#1}
\newcommand{\pgA}{Electromagnetism}
\newcommand{\pgB}{Fluid Mechanics}
\newcommand{\pgC}{Numerics}
\newcommand{\pgD}{Optimisation}

\newcommand{\csci}{Model Consistency, Topology and Geometry in Simulation and Design Optimisation}
\newcommand{\cscii}{Advanced Materials Modelling, Taking into Account Energy, Hysteresis and Other Losses}
\newcommand{\csciii}{Surrogate Models, Uncertainties, and Data-driven Simulations}
\newcommand{\csciv}{Integrated Thermal Design and Multiphysics}

\title{The CREATOR Project: Towards a Computational Electric Machine Laboratory\\
\small
Sebastian Schöps, Annette Muetze, Herbert De Gersem, Herbert Egger, Manfred Kaltenbacher and Markus Lazanowski%
}
\begin{document}
\maketitle

\begin{abstract} 
The Collaborative Research Centre TRR\,361/F90 CREATOR (2022--2030) aims at establishing a new paradigm for the simulation-driven design of electric machines. Increasing demands on efficiency, power density and sustainability require the integration of multiphysical effects, advanced materials and complex geometries into the design process. Traditional sequential workflows are no longer sufficient to address these challenges. CREATOR therefore combines expertise from electrical engineering, applied mathematics, fluid dynamics and materials science to establish integrated modelling, simulation and optimisation methodologies in a single large-scale project funded by the German and Austrian national funding agencies. This article provides an overview of the research vision, key achievements from the first funding period (2022--2026) and current developments towards a computational electric machine laboratory.
\end{abstract}

\section{Introduction}
\begin{figure}
    \centering
    \begin{tikzpicture}
        \begin{axis}[
            ybar,
            tudabarplot,
            bar width = 15pt,
            height = 6.0cm,
            width = 8.6cm,
            symbolic x coords={2015, 2016, 2017, 2018, 2019, 2020, 2021, 2022, 2023},
            xtick={2015, 2017, 2019, 2021, 2023},
            ytick={0,5,10,15},
            yticklabels={0\%, 5\%, 10\%, 15\%},
            tick label style={font=\sffamily\footnotesize},
            ymin=0, ymax=18,
            enlarge x limits=0.1,
            ymajorgrids=true,
            xlabel={Year},
            xlabel style={font=\sffamily\footnotesize},
            ylabel={Market share of elec. vehicles},
            ylabel style={font=\sffamily\footnotesize},
            legend style={
                at={(0.5,1.0)}, anchor=north, legend columns=-1,
                column sep=5pt,
            }
        ]
            \addplot+[color=TUDa-3a,fill=TUDa-3d] coordinates {
                (2015,  0.6)
                (2016,  0.9)
                (2017,  1.4)
                (2018,  2.2)
                (2019,  2.6)
                (2020,  4.2)
                (2021,  8.5)
                (2022, 13.6)
                (2023, 16.7)
            };
        \end{axis}
    \end{tikzpicture}
    \vspace*{-0.5em}
	\caption{Global market share of electric vehicles within passenger car sales \cite{Statista_EV_MarketShare}.}
    \label{fig:vehiclesales}
\end{figure}
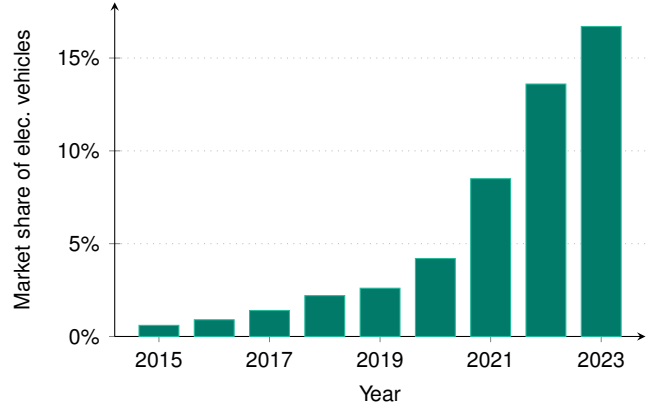

The transition to renewable energy sources is driving continuous developments in the electrification of industry and society. Replacing energy from fossil fuels with electricity from renewable sources is central to achieving climate goals. According to the International Energy Agency, electricity demand is expected to increase by over 70\,\% between 2010 and 2035 -- outpacing the growth of all other energy sources \cite[179]{IEA_2012aa}. 
Nearly 70\,\% of Europe’s electric energy is converted into mechanical energy by electric motors, making efficiency improvements highly impactful \cite{deAlmeida_2017aa}.
This pivotal role of electric motors has led to various international activities and regulations,
like the European Centre for Power Electronics as well as EU directives 2005/32/EG and 2023/1791, resulting in IEC 61800-9-2, EN 50598-2 and EN 50598-3 for energy-efficient electric machines and drives.
About 27\,\% of total US energy consumption in 2021 was used for transportation of people and goods
\cite[Table 2.1]{EIA_2025aa}. The electrification of transport is therefore of crucial importance for the transition to renewable energies. Over the last decade, electric vehicle (EV) sales have risen significantly (see \autoref{fig:vehiclesales}).
The European Commission's `Milestones of the Implementation Plan for Electrification in Road Transport’ targets a 60\,\% market share for EVs in new sales by 2030 \cite{Meyer_2017aa}.

Electro-mechanical energy conversion plays a key role in this transformation with important applications such as power generation, industrial manufacturing and electric mobility. By improving the performance of technical systems across these areas, innovations in electric machine design can make a significant contribution to increasing sustainability and reaching our climate goals.

Future motors demand higher power densities, minimised weight and costs, and reduced dependence on critical raw materials. Achieving these
goals requires a high level of customisation and adaptability to different operating conditions which drives traditional design workflows to their limits.

The consideration of complex machine topologies, new manufacturing techniques, transient operation regimes and thermal constraints calls for a novel fully
integrated simulation and design approach.
Addressing these challenges requires coordinated, long-term and interdisciplinary research efforts -- this is precisely the aim and structure of collaborative research centres in Germany and Austria.

This article presents the vision of the CREATOR project, highlights key achievements from the first funding period and outlines the framework in which further results are expected to enable the envisaged paradigm shift towards integrated, simulation-driven electric machine design. To this end, the article is structured as follows. Section~2 introduces the CREATOR project, its organisational framework and the interdisciplinary research challenges addressed within the collaborative research centre. Section~3 summarises the current state of the art in electric machine design and identifies key limitations of existing methodologies. Section~4 presents the overarching research vision and the four cross-sectional challenges that guide the scientific programme. Section~5 discusses recent developments and emerging trends that motivate the second funding period. Section~6 describes the interdisciplinary structure and collaboration mechanisms within the project. Section~7 highlights selected success stories from the first funding period, demonstrating the impact of the integrated research approach. Finally, Section~8 concludes the article and outlines future perspectives.

\section{The CREATOR Project}
Collaborative Research Centres (CRC) are long-term research programmes funded by the Deutsche Forschungsgemeinschaft (DFG) in Germany\footnote{\url{https://www.dfg.de/en/research-funding/funding-opportunities/programmes/coordinated-programmes/collaborative-research-centres}.} and/or the Österreichische Wissenschaftsfonds (FWF) in Austria\footnote{\url{https://www.fwf.ac.at/en/funding/portfolio/collaborations/special-research-areas}.}. They are designed to foster close collaboration across disciplines, enabling coordinated fundamental research over periods of up to eight years, typically involving 15--25 principal investigators and a total of around 50--100 researchers including doctoral candidates, postdocs and external experts with funding called Mercator Fellows by DFG. TRR\,361/F90 \emph{CREATOR}\footnote{\url{https://www.crc-creator.eu}.} (Computational electRic machinE lAboraTORy) is part of this programme as a transregional CRC (TRR) jointly funded by DFG and FWF, and represents the first collaborative research centre of this kind established across Germany and Austria. It is currently running in its second funding period, which started in March 2026 and runs until February 2030. It consists of about 20 (sub-)projects led by one or two principal investigators and its total funding for both periods will be around 20 million euros.
By joining interdisciplinary expertise from electrical engineering, numerical mathematics, fluid dynamics and materials science, CREATOR addresses the following cross-sectional challenges:\\[-1.5em]
\begin{itemize}\itemsep-0.2em
\item \csci.
\item \cscii.
\item \csciii.
\item \csciv.
\end{itemize}

The significance and viability of the research results are evaluated by comparison against benchmark problems and commercial software tools, as well as experimental data. Scientific goals are reported to and matched with the expectations from an industrial advisory board consisting of David Lowther (Infolytica, Siemens AG; McGill University), Daniel Scharfenstein (Robert Bosch GmbH) and Melanie Michon (Motor-CAD, Ansys).

In the second funding period, CREATOR advances its research activities on modelling, simulation and design methodologies for electric machines by considering multiphysical material behaviour, time-dependent operation conditions and advanced experimentation methodologies.
In addition, CREATOR will take up new research trends that have gained momentum in the last four years through the following activities:\\[-1.5em]
\begin{itemize}\itemsep-0.2em
    \item Consideration of novel cooling concepts and their impact on the design process.
    \item Incorporation of three-dimensional effects, e.g.\ in axial flux machines.
    \item Increased use of machine learning techniques in modelling, simulation and optimisation.
\end{itemize}

\begin{figure}
    \captionsetup[subfigure]{justification=centering}
    \centering
    \begin{subfigure}[b]{0.45\linewidth}
        \centering
        \includegraphics[height=2.2cm]{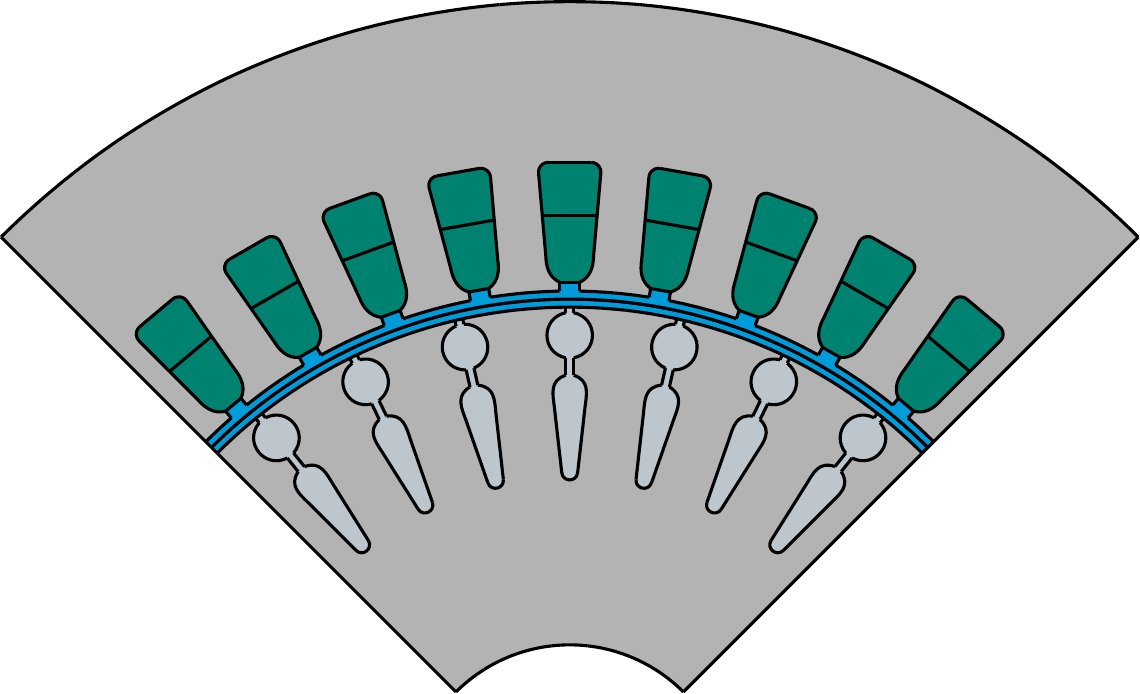}
        \caption{TUG IM}
        \label{fig:tugim}
    \end{subfigure}%
    \begin{subfigure}[b]{0.55\linewidth}
        \centering
        \includegraphics[height=2.2cm]{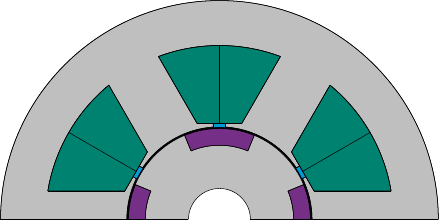}
        \caption{TUG PMSM}
        \label{fig:tugpmsm}
    \end{subfigure}%
    \caption{Two machine demonstrators from TU Graz~\cite{Muetze_2025aa}.}
\end{figure}

Furthermore, CREATOR strengthens its validation process through more sophisticated integrated experimental setups, which will deepen the understanding of the underlying physics and allow for better calibration of models and algorithms.
Common benchmark problems, validated simulation and experimental data are essential for the interdisciplinary collaboration between projects and also to ensure consistency and continuous progress in the research programme of the CRC. Examples include the TUG~IM (\autoref{fig:tugim}) and TUG~PMSM (\autoref{fig:tugpmsm}) machines, which are used as shared demonstrators for validating models and methods across different projects. Their design and measurement results have been made available recently in an open access data publication~\cite{Heidarikani_2024aa,Kumar-Dhakal_2024aa} as well as in a corresponding journal article~\cite{Muetze_2025aa}.

Through coordinated, interdisciplinary research, CREATOR aims to advance simulation and design methodologies for electric machines to a level where they can be reliably applied in industrial practice. By combining rigorous modelling, efficient numerical methods and data-driven techniques, these approaches will enable more systematic exploration of design spaces and support the development of high-performance machine concepts. At the same time, the close collaboration across disciplines fosters lasting integration of expertise and opens new perspectives beyond traditional research boundaries.

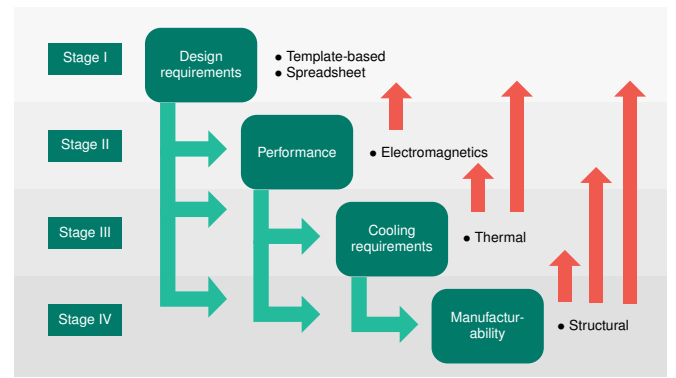
\begin{figure}
    \centering
    \scalebox{0.75}{%
        \begin{tikzpicture}[y=1pt, x=1pt, yscale=-1.1, xscale=1.1]
	\fill[color=black!3] (-43.9400,30.0200) rectangle (253.7000,73.7100);
	\fill[color=black!6] (-43.9400,73.7000) rectangle (253.7000,113.3900);
	\fill[color=black!9] (-43.9400,113.3900) rectangle (253.7000,153.0800);
	\fill[color=black!12] (-43.9400,153.0700) rectangle (253.7000,198.7600);
	\fill[TUDa-3d,text=white] (-28.3500,46.7700) rectangle (5.0000,60.9400) node[pos=.5,font=\sffamily\scriptsize] {Stage I};
	\fill[TUDa-3d,text=white] (-28.3500,86.4600) rectangle (5.0000,100.6300) node[pos=.5,font=\sffamily\scriptsize] {Stage II};
	\fill[TUDa-3d,text=white] (-28.3500,126.1400) rectangle (5.0000,140.3100) node[pos=.5,font=\sffamily\scriptsize] {Stage III};
	\fill[TUDa-3d,text=white] (-28.3500,165.8300) rectangle (5.0000,180.0000) node[pos=.5,font=\sffamily\scriptsize] {Stage IV};
	\fill[TUDa-3d,rounded corners=8,text=white] (15.5900,39.6800) rectangle (66.7800,73.7000) node[pos=.5,font=\sffamily\scriptsize,align=center] {Design\\requirements};
	\fill[TUDa-3d,rounded corners=8,text=white] (59.3400,79.3700) rectangle (110.5300,113.3900) node[pos=.5,font=\sffamily\scriptsize,align=center] {Performance};
	\fill[TUDa-3d,rounded corners=8,text=white] (102.6500,119.0500) rectangle (153.8300,153.0700) node[pos=.5,font=\sffamily\scriptsize,align=center] {Cooling\\requirements};
	\fill[TUDa-3d,rounded corners=8,text=white] (146.2900,158.7400) rectangle (197.4800,192.7600) node[pos=.5,font=\sffamily\scriptsize,align=center] {Manufactur-\\ability};
	\draw (100,57) node[font=\sffamily\scriptsize,align=left] {$\bullet$ Template-based\\ $\bullet$ Spreadsheet};
	\draw (145,97) node[font=\sffamily\scriptsize,align=left] {$\bullet$ Electromagnetics};
	\draw (175,135) node[font=\sffamily\scriptsize,align=left] {$\bullet$ Thermal};
	\draw (220,175) node[font=\sffamily\scriptsize,align=left] {$\bullet$ Structural};
	\fill[color=TUDa-9a] (127.2100,70.8700) rectangle (133.2100,86.2900);
	\fill[color=TUDa-9a] (122.9200,71.8400) -- (137.5000,71.8400) -- (130.2100,64.5600) -- cycle;
	\fill[color=TUDa-9a] (164.5600,107.5900) rectangle (170.5600,123.7800);
	\fill[color=TUDa-9a] (160.2700,108.5700) -- (174.8500,108.5700) -- (167.5700,101.2800) -- cycle;
	\fill[color=TUDa-9a] (182.2300,70.0900) rectangle (188.2300,123.7900);
	\fill[color=TUDa-9a] (177.9400,71.0700) -- (192.5200,71.0700) -- (185.2300,63.7800) -- cycle;
	\fill[color=TUDa-9a] (218.3600,109.4500) rectangle (224.3600,165.1500);
	\fill[color=TUDa-9a] (214.0700,110.4300) -- (228.6500,110.4300) -- (221.3600,103.1400) -- cycle;
	\fill[color=TUDa-9a] (204.1100,147.2800) rectangle (210.1100,164.7000);
	\fill[color=TUDa-9a] (199.8200,148.2600) -- (214.4000,148.2600) -- (207.1100,140.9700) -- cycle;
	\fill[color=TUDa-9a] (233.8000,70.0900) rectangle (239.8000,165.5500);
	\fill[color=TUDa-9a] (229.5100,71.0700) -- (244.0900,71.0700) -- (236.8000,63.7800) -- cycle;
	\fill[color=TUDa-3a] (45.6300,98.4200) -- (22.4700,98.4200) -- (22.4700,73.8300) -- (29.4700,73.8300) -- (29.4700,91.4200) -- (45.6300,91.4200) -- cycle;
	\fill[color=TUDa-3a] (44.4900,103.4200) -- (44.4900,86.4100) -- (52.9900,94.9200) -- cycle;
	\fill[color=TUDa-3a] (45.6300,125.9700) -- (22.4700,125.9700) -- (22.4700,94.9200) -- (29.4700,94.9200) -- (29.4700,118.9700) -- (45.6300,118.9700) -- cycle;
	\fill[color=TUDa-3a] (44.4900,130.9700) -- (44.4900,113.9700) -- (52.9900,122.4700) -- cycle;
	\fill[color=TUDa-3a] (45.6300,166.8800) -- (22.4700,166.8800) -- (22.4700,122.4700) -- (29.4700,122.4700) -- (29.4700,159.8800) -- (45.6300,159.8800) -- cycle;
	\fill[color=TUDa-3a] (44.4900,171.8900) -- (44.4900,152.8800) -- (52.9900,163.3800) -- cycle;
	\fill[color=TUDa-3a] (88.1900,138.2600) -- (65.0300,138.2600) -- (65.0300,113.6800) -- (72.0300,113.6800) -- (72.0300,131.2600) -- (88.1900,131.2600) -- cycle;
	\fill[color=TUDa-3a] (87.0500,143.2700) -- (87.0500,126.2600) -- (95.5600,134.7600) -- cycle;
	\fill[color=TUDa-3a] (88.1900,173.8200) -- (65.0300,173.8200) -- (65.0300,136.7600) -- (72.0300,136.7600) -- (72.0300,166.8200) -- (88.1900,166.8200) -- cycle;
	\fill[color=TUDa-3a] (87.0500,178.8200) -- (87.0500,161.8100) -- (95.5600,170.3200) -- cycle;
	\fill[color=TUDa-3a] (132.8900,178.4300) -- (109.7300,178.4300) -- (109.7300,153.3700) -- (116.7300,153.3700) -- (116.7300,171.4300) -- (132.8900,171.4300) -- cycle;
	\fill[color=TUDa-3a] (131.7500,183.4300) -- (131.7500,166.4200) -- (140.2500,174.9200) -- cycle;
\end{tikzpicture}
    }
    \caption{Current design workflow, illustration based on Rosu et al. \cite{Rosu_2017aa}.}
    \label{fig:multi_workflow}
\end{figure}%

\section{State of the Art}
Today’s electric machine design workflows are rather sequential (see \autoref{fig:multi_workflow}): analytic formulas and tabulated data guide initial stages like sizing and fundamental design choices (materials, topologies). Specifications include speed, power, torque, efficiency, thermal properties, weight and size for a few operating points. More time-consuming 2D finite-element simulations and parametric optimisations run on high-performance clusters. Even more complex 3D simulation models are used to further refine designs.
Tabulated data and equivalent circuit models support material and thermal behaviour analysis. Thermal and acoustic limits are then usually verified in computationally intensive post-processing routines. Multiple iterations may be required
to integrate magnetic, thermal and acoustic design.
Besides proprietary software like 
Altair Flux (now Siemens) \cite{altair_flux},
ANSYS Maxwell 2D/3D \cite{ansys_maxwell}, Motor-CAD \cite{ansys_motorcad} and RMxprt \cite{ansys_rmxprt}, 
COMSOL \cite{comsol}, 
CST EM Studio \cite{cst}, 
FEmag \cite{femag}, 
JMAG \cite{jmag}, 
LCM SyMSpace \cite{symspace}, 
Siemens Simcenter MagNet (formerly Infolytica MagNet) \cite{simcenter_magnet} and CCM+ \cite{simcenter_ccm}, 
in-house codes and free tools like FEMM \cite{femm}, NGSolve \cite{ngsolve} or Onelab/GetDP \cite{getdp} are widely used in academia and industry.

\subsection{Machine Designs}
Modern off-the-shelf induction motors for industrial applications achieve efficiencies from 70--80\,\% at 1\,kW, to 90\,\% and higher at 90\,kW rated power~\cite{Mecrow_2008aa}. However, efficiency may drop significantly in variable-speed scenarios. For example, an electric traction motor with peak efficiency above 97\,\% typically averages around 92\,\% in drive cycle tests like the Worldwide Harmonised Light Vehicles Test Procedure (WLTP, \cite{UNECE_WLTP_154}), sometimes falling below 90\,\%, notably at low or high speeds. 
Besides efficiency, there is a higher demand on the power density. Bosch’s motor platform\footnote{\url{https://www.bosch-engineering.com/services/mobility-solutions/motion/powertrain-systems/}} increased its power density from 12 kW/L to 40 kW/L from 2012 to 2024, a more than 200\,\% rise in power density, while production costs in the same time frame were reduced by a factor of three.
Such highly optimised machine designs define the current state of the art for electric traction applications.

\subsection{New Opportunities}
Technological progress in power electronics, novel materials, advanced manufacturing techniques, innovative cooling concepts and control strategies have opened up a range of possibilities for new innovations, principally allowing a further increase of power and torque densities and energy conversion efficiency. According to the start-up company Additive Drives \cite{AdditiveDrives_2025}
\begin{quote}
\textit{better cooling, better winding topologies, new insulation material and ultimately many design freedoms} [are] \textit{enabled by 3D printing.} [These] \textit{achieve up to 45\,\% higher power density}
[in comparison with conventional designs].
\end{quote}
However, such innovations must still be fully integrated into future machine design workflows. 
In March 2024, the US Department of Energy released the newest version of their `Electric Drive Technical Team Roadmap', in which future performance targets of drive systems and components
are defined, see \autoref{fig:motor_targets} and \cite{USDrive_2024aa}.
According to this document
\begin{quote}
... \emph{systems must continue to decrease cost to achieve cost of ownership equal to or less than that of ICEs} {\normalfont [internal combustion engines]} vehicles. 
{\normalfont [...]} \emph{The efficiency of the overall drive cycle needs to be improved to reduce the amount of energy storage required on the vehicle.}{\normalfont [Achieving these ambitious targets requires]}
\emph{an overall systems and ecosystem view to address and resolve the heterogeneous or multi-physics integration of materials, nano-carbon infused metals, a new class of isolation materials, high-temperature materials and new thermal management techniques.}
\end{quote}
The consideration of these aspects clearly exceeds the capabilities of current computational workflows. It requires a concerted effort of theoretical advances in different scientific research fields and a novel integrated multiphysical simulation, optimisation and design approach.

\begin{table}
    \caption{Power electronics and electric motor targets (US DoE \cite{USDrive_2024aa}).}
    \centering
    \scriptsize
    \sffamily
    \arrayrulecolor{white}
    \renewcommand{\arraystretch}{1.4}
    \begin{tabular}{l|c|c|c}
        \rowcolor{TUDa-3d}
            \multicolumn{4}{c}{\textbf{\color{white}Power Electronics Targets}}
        \\\hline
            \rowcolor{TUDa-3d!30}
            \cellcolor{TUDa-3d} \textbf{\color{white}Year} & 2025 & 2030 & 2035
        \\\hline
            \rowcolor{TUDa-3d!10}
            \cellcolor{TUDa-3d} \textbf{\color{white}Power Level (kW, Peak)} & 150  & 200  & 225
        \\\hline
        \rowcolor{TUDa-3d!30}
            \cellcolor{TUDa-3d} \textbf{\color{white}Voltage (V)} & 600  & 800  & 800
        \\\hline
        \rowcolor{TUDa-3d!10}
            \cellcolor{TUDa-3d} \textbf{\color{white}Cost (\$/kW)} & 1.80  & 1.35  & 1.20
        \\\hline
        \rowcolor{TUDa-3d!30}
            \cellcolor{TUDa-3d} \textbf{\color{white}Power Density (kW/L)} & 150  & 200  & 225
    \end{tabular}
    ~\\[0.5em]
    \begin{tabular}{l|c|c|c}
        \rowcolor{TUDa-3d}
            \multicolumn{4}{c}{\textbf{\color{white}Electric Motor Targets}}
        \\\hline
            \rowcolor{TUDa-3d!30}
            \cellcolor{TUDa-3d} \textbf{\color{white}Year} & 2025 & 2030 & 2035
        \\\hline
            \rowcolor{TUDa-3d!10}
            \cellcolor{TUDa-3d} \textbf{\color{white}Power Level (kW, Peak)} & 150  & 200  & 225
        \\\hline
        \rowcolor{TUDa-3d!30}
            \cellcolor{TUDa-3d} \textbf{\color{white}Voltage (V)} & 600  & 800  & 800
        \\\hline
        \rowcolor{TUDa-3d!10}
            \cellcolor{TUDa-3d} \textbf{\color{white}Cost (\$/kW)} & 2.20  & 1.65  & 1.47
        \\\hline
        \rowcolor{TUDa-3d!30}
            \cellcolor{TUDa-3d} \textbf{\color{white}Power Density (kW/L)} & 75  & 100  & 112.5
    \end{tabular}
    \arrayrulecolor{black}
    \label{fig:motor_targets}
\end{table}

\section{Research Vision}
CREATOR brings together a multidisciplinary team of experts in electrical engineering, fluid dynamics, materials science, numerical mathematics
and optimisation, with the shared vision to transform the foundations of electric machine design through
simulation-driven innovation. 
The goal is to develop novel simulation and optimisation tools capable of addressing the complex multiphysics simulations and high-dimensional optimisation problems under uncertainty that are required to unleash the full potential of next-generation machine designs. 
Based on a strong history of collaborative research and industrial partnerships, the project strives to create design methodologies that are not only grounded in rigorous theoretical analysis but also efficient, robust and adaptable to the evolving demands of industrial practice. 
By pushing the boundaries of current design workflows, the aim is to provide tools for novel, high-performance and innovation-ready electric machine technologies. 
Intensive collaboration strengthens the existing bonds between the involved research areas, enabling scientific progress across the boundaries of traditional research disciplines.

CREATOR groups the research questions into four interdisciplinary cross-sectional challenges that serve as a strategic framework and a guideline to align the individual work programmes towards an overarching research vision.

\subsection{\csci~(CSC~1)} 
CSC~1 tackles the challenge of flexible representation and automated handling of model parameters, geometry and topology -- key to seamless, computer-aided design. 
Precision in feature handling and the integration of multiple moving domains are essential for efficient simulation across space and time.
Geometric and topological variations stem from design optimisation, uncertainty quantification and multiphysical interactions such as heat, mechanical stress and acoustics.

These challenges raise several research questions:
How can models remain consistent across all
design stages, computational tools and real-world operation?
Which formulations and numerical
methods allow ultra-fast evaluations, given that companies simulate 100,000 designs weekly -- each requiring hundreds of finite element calculations across rotor angles and excitations?
Which effects must be captured on the finite element level, and what can be simplified as boundary conditions or post-processing?
\begin{figure}
    \centering
    \includegraphics[width=\linewidth,height=4.5cm]{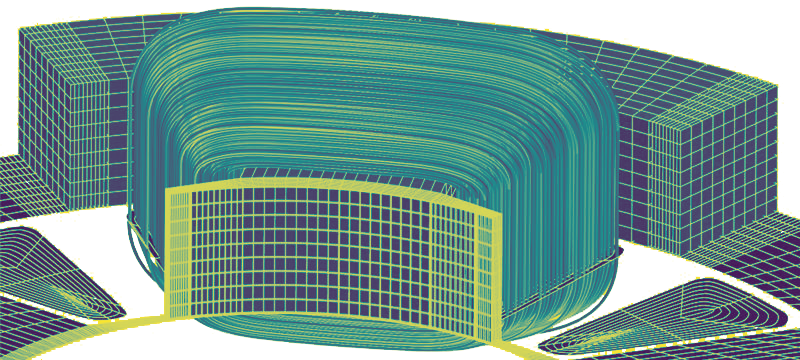}
    \caption{Three-dimensional volumetric spline model with resolved windings.}
    \label{fig:csc1}
\end{figure}  
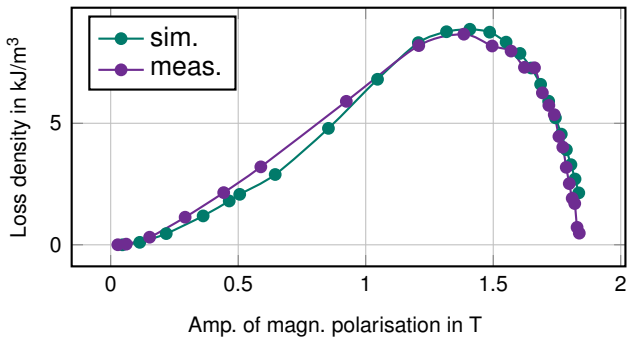
\begin{figure}
    \centering
\begin{tikzpicture}[font=\sffamily]
\begin{axis}[
    width=\linewidth,
    height=5cm,
    tick label style={font=\sffamily\footnotesize},
    xlabel style={font=\sffamily\footnotesize},
    ylabel style={font=\sffamily\footnotesize},
    ylabel={Loss density 
        in kJ/m$^{\text{3}}$},
    ylabel style={yshift=-1.5em},
    ytick={0,5,10}, 
    yticklabels={0,5,10}, 
    xlabel={Amp. of magn. polarisation 
        in T},
    xtick={0,0.5,1,1.5,2}, 
    xticklabels={0,0.5,1,1.5,2}, 
    xlabel style={xshift=-0.5em},
    grid=major,
    legend pos=north west,
    legend cell align={left},
    thick,
]
\addplot[
    color=TUDa-3d,
    mark=*,
    smooth
] table [
    x index=0,
    y expr=\thisrowno{1}/1000,
    col sep=comma,
    header=false
] {figures/rot_losses_jmmm.csv};
\addplot[
    color=TUDa-11b,
    mark=*,
    smooth
] table [
    x index=2,
    y expr=\thisrowno{3}/1000,
    col sep=comma,
    header=false
] {figures/rot_losses_jmmm.csv};
\addlegendentry{sim.}
\addlegendentry{meas.}
\end{axis}
\end{tikzpicture}
    \vspace{-0.5em}
    \caption{Rotational hysteresis losses, see for details~\cite{Sauseng_2024aa}.}
    \label{fig:csc2}
\end{figure}
A result from the first funding period is shown in \autoref{fig:csc1}. It combines expertise on electric machines and volumetric spline modelling from several projects, which are introduced in \autoref{sec:interdisciplinarity}.

\subsection{\cscii~(CSC~2)} 
CSC~2 tackles the complex challenge of refining thermal loss modelling -- capturing hysteresis, eddy currents and thermal dependencies. This research is crucial for miniaturisation, next-generation composite and printed materials as well as a deeper understanding of transient machine behaviour.
Therefore, energy balances must be analysed across all relevant length and time scales, integrating electronic, magnetic, mechanical and thermal effects in simulation and optimisation.

Key research questions include: Which physical effects, such as hysteresis, heat and stress, possibly on different time-scales, significantly impact electric machine performance? Loss prediction is also critical. 
Finding effective material models is crucial. How can their parameters be precisely determined from experiments or simulations to enable next-generation material models?

\autoref{fig:csc2} showcases a result from the first funding period on hysteresis loss modelling and measurement based on a novel experimental setup developed in Graz.

\subsection{\csciii~(CSC~3)} 
CSC~3 addresses the need for well-designed experiments to validate models and methods.
Model parameters must often be identified through estimation or data-driven methods.
Phenomenological, surrogate and equivalent-circuit models require careful calibration via inverse problem formulations and machine learning.
Uncertainties must be minimised through optimised experiments and product design.

Exemplary research questions are: How can the predictive accuracy of models be improved using additional data? Which methodological approach -- such as data-driven modelling, response surface techniques or model order reduction -- is best suited for specific applications?
\begin{figure}
    \centering
    \input{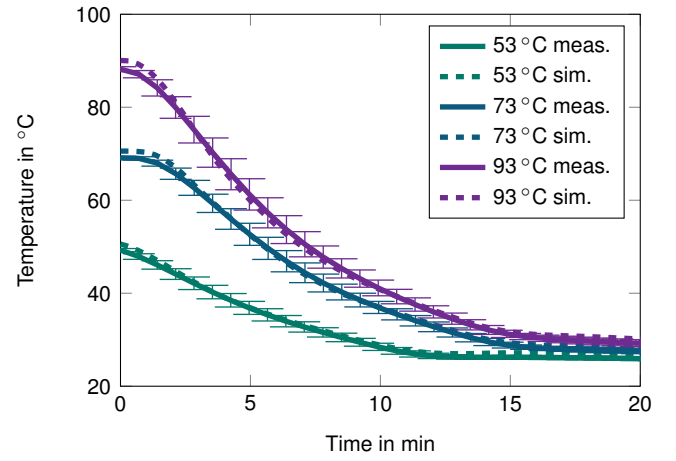}
    \caption{Comparison of simulation and measurements at three positions within the winding for TUG IM \cite{Bergfried_2024ac}.}
    \label{fig:csc3}
\end{figure}

A result from the first funding period is shown in \autoref{fig:csc3}: the validation of a thermal machine model using experimental data.
\subsection{\csciv~(CSC~4)}
CSC~4 specifically focuses on cooling and mechanics, key for the next-generation of electric machines.
This important topic has been introduced as a fourth dedicated challenge for the second funding period,
as reducing volume to achieve the ambitious electric motor targets also decreases heat capacity. Advanced cooling techniques and a precise physical understanding of thermal behaviour are more critical than ever.

This raises several key research questions: What are the exact physical mechanisms governing different cooling technologies, including heat transfer and fluid dynamics? How accurately can -- and should -- these be modelled?
And how can the corresponding models be integrated into the design workflow for electric machines?
\begin{figure}    
    \centering
    \scalebox{1.1}{
    \begin{tikzpicture}[font=\tiny\sffamily,inner sep = 0cm]
        \node[anchor=south west, inner sep=0] (image) at (0,0) {
            \includegraphics[width=7.2cm]{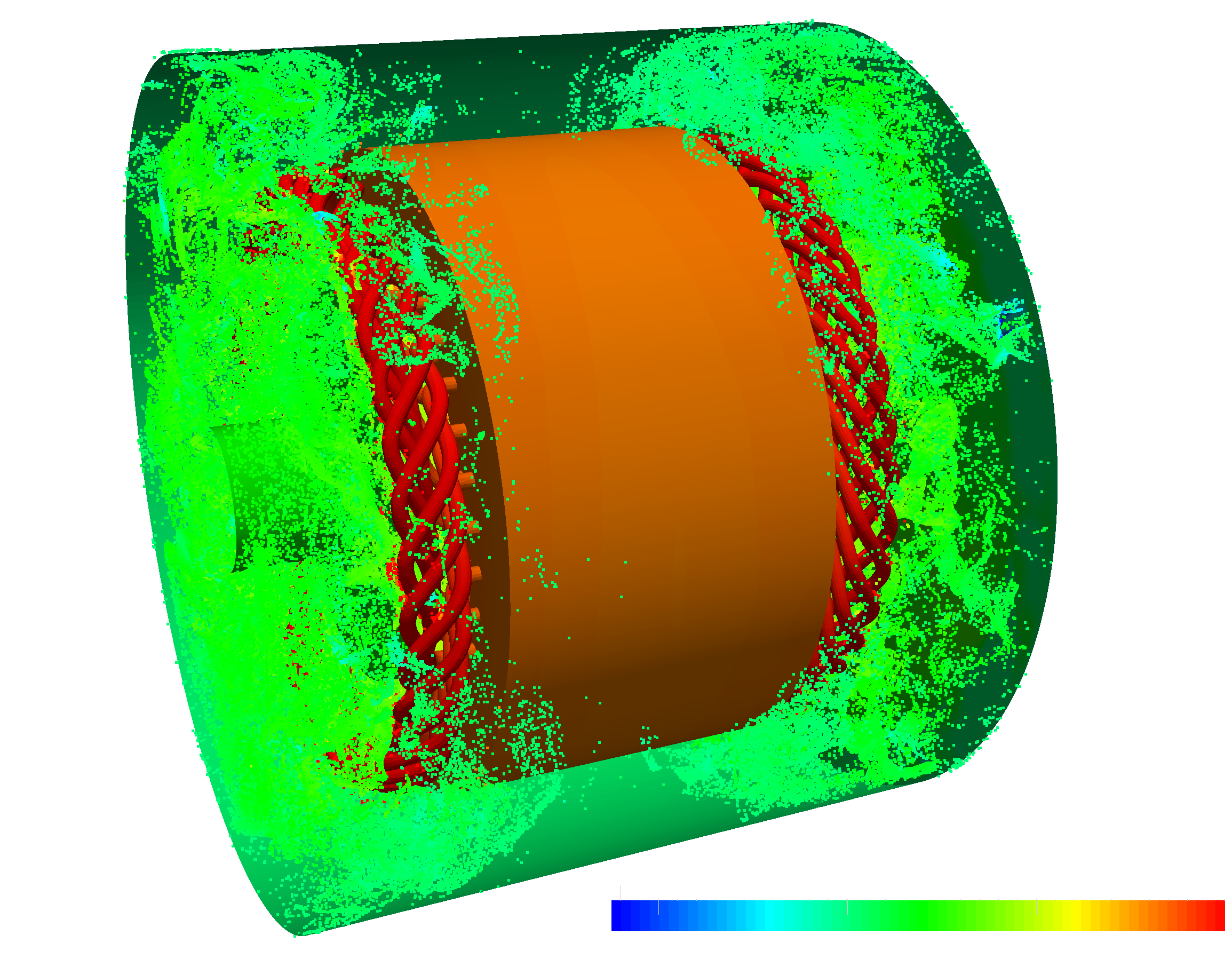}
        };
        \node[anchor=south] at (5.5cm,0.5cm) {Temperature};
        \node[anchor=south] at (3.6cm,0cm) {275\;K};
        \node[anchor=south] at (7.0cm,0cm) {340\;K};
    \end{tikzpicture}
    }
    \caption{Multiphysics simulation results of retrofitted spray cooling as shown in \autoref{fig:rjic} for TUG IM, see \cite{Lohner_2025aa}.}
    \label{fig:csc4}
\end{figure}
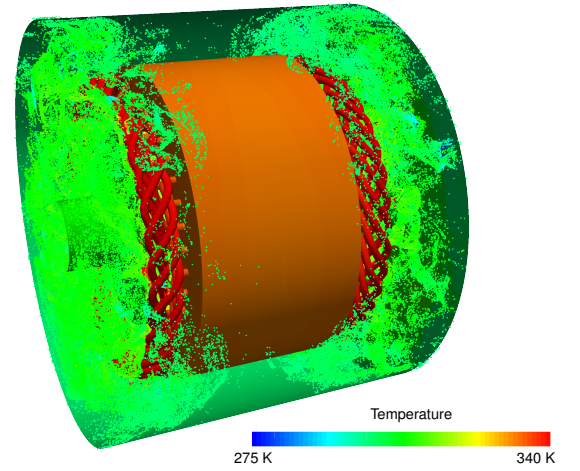

A multiphysical simulation result from the first funding period related to fluid cooling, now considered part of CSC~4, is presented in \autoref{fig:csc4}. This is joint work of several groups from CREATOR and the Center for Computational Fluid Dynamics from George Mason University.

All research projects and members of CREATOR address aspects of these cross-sectional research challenges and contribute to the solution of the underlying research questions.

\section{Recent Developments and Emerging Trends}
Over the past three years, several research trends have gained momentum and will shape the next phase of innovation. 
We identified several topics of particular relevance for the design workflow of electric machines.

\subsection{Cooling of Electric Machines}
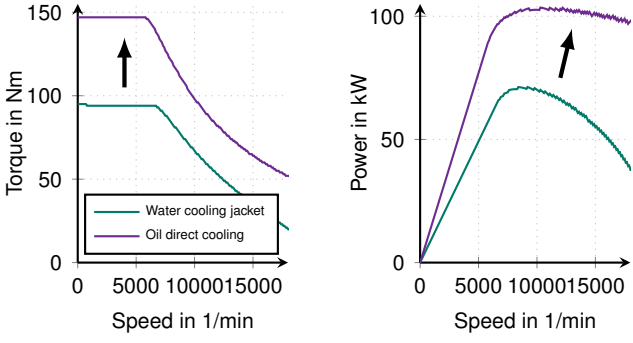
\begin{figure}[t]
    \centering
    \captionsetup[subfigure]{justification=centering}
    \begin{subfigure}{0.49\linewidth} 
        \centering
        \begin{tikzpicture}[font=\sffamily]
            \begin{axis}[
                tudalineplot,
                legend style={font=\sffamily\tiny},
                tick label style={font=\sffamily\footnotesize},
                xlabel style={font=\sffamily\footnotesize},
                ylabel style={font=\sffamily\footnotesize, yshift=-1.2em},
                scaled x ticks = false,
                xtick = {0,5000,10000,15000,20000},
                xticklabels = {0,5000,10000,15000,20000},
                ytick = {0,50,100,150},
                yticklabels = {0,50,100,150},
                xlabel={Speed in 1/min},
                xmin={0},
                xmax={18000},
                ymin={0},
                ymax={155},
                ylabel={Torque in Nm},
                width=\linewidth,
                height=5cm,
                grid=major,
                legend pos=south west,
            ]
            \addplot +[color=TUDa-3d,mark=none] table [x=n2, y=M2, col sep=comma] {figures/yves_cooling.csv};
            \addplot +[color=TUDa-11b,mark=none] table [x=n3, y=M3, col sep=comma] {figures/yves_cooling.csv};
            \draw[ultra thick, -latex, black] (axis cs:4e3,105) -- (axis cs:4e3,135);
            \legend{Water cooling jacket,Oil direct cooling}
            \end{axis}
        \end{tikzpicture}
        \caption{Simulation results: torque as a function of speed}
        \label{fig:torque_power}
    \end{subfigure}%
    \hfill
    \begin{subfigure}{0.49\linewidth} 
        \centering
        \begin{tikzpicture}[font=\sffamily]
            \begin{axis}[
                tudalineplot,
                legend style={font=\sffamily\footnotesize},
                tick label style={font=\sffamily\footnotesize},
                xlabel style={font=\sffamily\footnotesize},
                ylabel style={font=\sffamily\footnotesize, yshift=-1.2em},
                scaled x ticks = false,
                xtick = {0,5000,10000,15000,20000},
                xticklabels = {0,5000,10000,15000,20000},
                ytick = {0,50,100},
                yticklabels = {0,50,100},
                xlabel={Speed in 1/min},
                xmin={0},
                xmax={18000},
                ymin={0},
                ymax={105},
                ylabel={Power in kW},
                width=\linewidth,
                height=5cm,
                grid=major
            ]
            \addplot +[color=TUDa-3d,mark=none] table [x=n2, y=Pmech2, col sep=comma] {figures/yves_cooling.csv};
            \addplot +[color=TUDa-11b,mark=none] table [x=n3, y=Pmech3, col sep=comma] {figures/yves_cooling.csv};
            \draw[ultra thick, -latex, black] (axis cs:1.2e4,75) -- (axis cs:1.3e4,95);
            \end{axis}
        \end{tikzpicture}
        \caption{Simulation results: power as a function of speed}
        \label{fig:torque_speed}
    \end{subfigure}
    \caption{Comparison of cooling concepts: water-jacket vs. direct oil cooling.
    }
    \label{fig:cooling_comparison}
\end{figure}

On the application side, the challenge of cooling electric machines has become increasingly central. Achieving the desired power and torque densities requires efficient removal of loss-induced heat. Established cooling strategies, such as housing-integrated water jackets, are reaching their performance limits, necessitating the development of advanced direct liquid, spray-based and phase-change cooling techniques. These novel methods are even more crucial where space is limited but operational reliability is required, for example in electric mobility applications.
Enhanced cooling capabilities allow for a higher current that can be applied on a given torque producing surface within a machine.
This in turn increases power and torque density as well as efficiency, since elevated temperatures would otherwise increase joules losses in the winding. 
\autoref{fig:cooling_comparison} illustrates the potential of novel cooling concepts based on the use of oil as a cooling medium. Further cooling concepts investigated in CREATOR are depicted in \autoref{fig:cooling_types}. 
A detailed evaluation of advanced cooling methods concerning achievable thermal conductivities and thermally admissible conductor current densities related to the machine's developed torque and power has been given in \cite{Shams-Ghahfarokhi_2023aa}. 
According to these findings, direct liquid cooling (e.g. oil jet cooling) allows for increasing power density by a factor of about 1.5 compared to indirect liquid cooling (e.g. water jacket cooling). Alternatively, the original torque or power could also be realised with a motor-size reduction of 20--40\,\%.
Novel cooling methods therefore are crucial for high-performance applications.

Research in the new CSC4 has been strengthened by the introduction of two new projects, related to stator spray and radial rotor spray cooling, and the inclusion of a new Mercator Fellow. A demonstrator including spray and heat-pipe cooling has been designed and is currently being built in Darmstadt.

\subsection{Three-dimensional Effects}
Driven by the demand for increased energy efficiency and power density, the accurate consideration of three-dimensional effects has become even more relevant. In conventional radial flux machines, these effects include, for example, AC losses in end windings and eddy current losses in permanent magnets. Furthermore, axial flux machines (AFMs) have attracted increasing interest, exemplified by the acquisition of YASA Ltd. by Mercedes-Benz AG. To accurately capture such effects and to optimise the AFM performance, computationally efficient three-dimensional simulation workflows are required. A strategic partnership with the European MAXIMA (Modular AXIal flux Motor for Automotive) project has been established \footnote{See \url{https://maxima-he.eu/}.}.

The existing projects \project{A03}, \project{C04} and \project{D02} are extending their methodology and evaluations towards pronounced three-dimensional phenomena, in particular AFMs, the new \project{C06} (succeeding \project{C01}) and new Mercator Fellow have been added to address this in CSC1.

\subsection{Machine Learning}
On the computational side, the widespread use of machine learning (ML) has driven a paradigm shift towards data-driven techniques. Recent advancements in surrogate modelling, reduced-order models and neural network-based methods have opened new possibilities for accelerating computations, improving predictive accuracy and handling uncertainty. For example, we have identified a need for research regarding the integration of physics, in particular electric machine know-how, into the learning process, i.e. hybrid modelling approaches that combine data-driven and physics-based techniques, rather than the development of new network architectures or the naïve application of neural networks. Either way, measurements and high-fidelity numerical simulations with guaranteed accuracy remain indispensable since learning techniques require data, in many cases even large-scale datasets.

Data-driven methods and ML were already featured in several projects during the first funding period. A team from CREATOR received the first prize in the interpolation category of the Galileo contest using a model based on polynomial chaos Kriging and Gaussian process regression~\cite{Solimene_2025aa}. To further support the research questions related to CSC~3, ML has been given increased emphasis within the research agenda of the second funding period. Additionally, a new project has been introduced focusing on learning port-Hamiltonian machine models. The new projects and the adaptations strengthen the CRC's capacity to reach our overarching goals of enhancing predictive accuracy, computational efficiency and practical applicability, bridging the gap between fundamental research and industrial implementation.

\section{Interdisciplinarity}\label{sec:interdisciplinarity}
As an interdisciplinary collaborative research project, CREATOR requires a high level of domain-specific scientific expertise as well as a set of coordinated measures to foster interdisciplinary cooperation and synergies. In the following, we provide a brief overview of the main structure, organisation, collaborative measures and management of the CRC.

Each research project is led by principal investigators representing the relevant scientific expertise. Based on their primary disciplinary focus, the projects are assigned to one of the four project areas:
{A: \pgA}, {B: \pgB}, {C: \pgC} or {D: \pgD}. Each project is expected to perform specialised research in its area while also contributing to the interdisciplinary goals of the CRC by addressing at least one of the cross-sectional challenges mentioned above.

\subsection{\pgA}
Area A focuses on advancing electromagnetic and thermal modelling and simulation methodologies with an emphasis on efficiency, accuracy and reliability.
\textbf{A01}
investigates the confidence levels of performance-map-based drive cycle analysis. Downscaling of drive cycles to laboratory level is considered, and the importance of efficiency map resolution is evaluated and compared with high-fidelity simulations. The project provides two complete datasets for electric motors, which are now used for validation across projects of the CRC.
\textbf{A02}
develops an isogeometric domain decomposition solver for magnetic field simulations using NURBS geometry representations to improve meshing and accuracy. Mortaring across non-matching domain interfaces, e.g. rotor and stator, and surrogate modelling based on model order reduction and machine learning are considered to enable fast and accurate machine simulation.
\textbf{A03}
focuses on adaptive-resolution machine models, addressing skewing effects and eccentricity by extended 2D simulations. A multi-fidelity approach is established, integrating machine learning-based surrogate models and equivalent circuit representations to enhance computational efficiency. A novel multi-patch PINN approach is developed to address material discontinuities.
\textbf{A04}
considers electric and thermal stresses in machine insulation, providing critical insights into insulation breakdown. A coupled electrothermal simulation framework is developed allowing for extensive validation and calibration to measurement data.
The potential of novel cooling concepts for
high-power-density motors is demonstrated by
analysis of an induction machine.
\textbf{A05}
advances the multiscale simulation of hysteresis and eddy-current losses in magnetic core materials using computational homogenisation. Machine learning hysterons are used for micro-to-macro scale bridging, enabling accurate analysis of microstructural effects on material performance.
The developed methodology is validated for two Fe-Si-B microstructure samples.
\textbf{A06} develops a new electric machine demonstrator for battery electric vehicles, benchmarks commercial software tools against novel CREATOR-developed methodologies and investigates high-frequency effects due to power electronics.

By integrating advanced field simulations, material modelling and drive cycle assessments, the research efforts of project area A significantly enhance the predictive power and efficiency of computational machine models required for the development of high-performance and sustainable electric drives.

\subsection{\pgB}
This area focuses on the investigation of advanced cooling techniques for electric motors. Two complementary multiphase cooling strategies are explored numerically and experimentally, offering significantly improved heat transfer rates compared to established methodologies.
\textbf{B02} studies the potential of heat pipes for rotor cooling in electric motors. An extended Discontinuous Galerkin method is developed to simulate liquid--vapour flows and energy transfer in such devices. Key innovations include a parametrised level-set method to improve numerical stability and allow for larger time steps in two-phase simulations in the presence of strong capillary effects.
\textbf{B03} studies spray-based rotor cooling. Droplet impact and spray characteristics are captured by high-speed recordings, while instantaneous temperature and heat flux at the cylinder surface are measured experimentally and determined theoretically. An empirical spray-cooling model based on the Stanton number is developed, leading to improved heat flux predictions in thermal simulations.
\textbf{B04} investigates compound drop-based cooling to enhance thermal management of electric machines, particularly focusing on end windings.
\textbf{B05} performs an experimental study of heat transfer processes in the air gap and at the end windings of an oil-spray-cooled electric motor using advanced optical measurement techniques.

The advances in project area B demonstrate the potential of novel cooling strategies to enhance heat transfer and thereby improve motor performance and operating regimes.

\subsection{\pgC}
Area C focuses on advancing theoretical foundations and computational methods for the analysis and simulation of electric machines, including multiscale and multiphysical aspects.
\textbf{C01}
develops a multiphysical simulation framework integrating isogeometric analysis and the new Magnetic Oriented Node Analysis \cite{Cortes-Garcia_2022ab} to improve coupled field-circuit simulations. Co-simulation approaches are investigated for electromagnetic--thermal coupling, and parallelisation techniques are developed to enhance simulation efficiency and accuracy for complex electric machine models.
\textbf{C02}
focuses on energy-consistent modelling and structure-preserving discretisation, incorporating advanced material models for magnetic hysteresis, anisotropy and eddy current losses. It further develops reduced-order models for magnetic fields, torque and heat transfer, ensuring mathematically well-posed formulations validated against experimental data and proprietary software.
\textbf{C03}
develops an isogeometric analysis framework for electric machine design, allowing the analysis of complex geometries. It utilises local refinement based on multi-level B\'ezier extraction and level sets for trimming. Open-source benchmarks for cut element integration are developed, and conditions to improve curvature integrability of watertight splines are derived.
\textbf{C04}
explores the potential of novel space-time finite element methods for electric machine analysis, focusing on improving the accuracy and efficiency of transient simulations of magnetic and thermal fields in rotating machines. An adaptive least-squares framework is introduced for systematic error estimation and multiphysics coupling.
\textbf{C05}
studies space-time boundary element formulations for noise and vibration analysis of electric drives. The multivariate Adaptive Cross Approximation reduces computing time and improves storage to enable three-dimensional computation of sound scattering of electrical machines in the time domain. The generalised convolution quadrature method is further improved.
\textbf{C06} advances multiphysics modelling using isogeometric analysis, improving the integration of electromagnetic, structural and thermal simulations with a specific focus on axial flux machines.

The research efforts of project area C lead to significant advancements in modelling, analysis and numerical methods, improving reliability, accuracy and time-to-solution of electric machine simulations.

\subsection{\pgD}
The last area D studies optimisation techniques for electric machine design, covering surrogate modelling, topology optimisation, uncertainty quantification and parameter identification.
\textbf{D01}
investigates data-driven surrogate models for simulation and uncertainty quantification for electric machines. Machine learning techniques, such as Gaussian Process Regression and Neural Networks, are integrated into the optimisation workflow, enabling rapid evaluation of design alternatives, efficient sensitivity analyses, extensive parametric studies and uncertainty quantification.
\textbf{D02}
focuses on multiphysics topology optimisation for electric machines under mechanical and manufacturability constraints. Topological derivatives, a level-set-based approach and multimaterial topology optimisation algorithms are developed to integrate electromagnetic and thermal performance metrics. Transient settings and full drive cycles are considered.
\textbf{D03}
develops efficient methods for robust shape and topology optimisation addressing uncertainties in material properties and operating conditions. A bilevel optimisation framework is employed and a novel robust topological derivative is proposed and applied to electric machine design. Optimal experimental design is considered for identification of magnetic material parameters.
\textbf{D04}
focuses on determining local magnetic properties and their impact on electric machine performance. Experimental setups for measuring local magnetic material properties, including eddy currents and hysteresis, are developed. A physically consistent vector hysteresis operator is introduced, allowing the systematic incorporation of material data into high-fidelity simulations.
\textbf{D06} investigates port-Hamiltonian neural networks (PHNNs) for efficient surrogate modelling and uncertainty quantification of electric machines, ensuring energy-consistent reduced-order models.

The research efforts of project area D significantly advance optimisation methodologies for electric machine design, enabling efficient handling of high-dimensional design spaces and uncertainties. These developments provide a key foundation for integrating simulation, data-driven modelling and optimisation into a unified design framework.

\subsection{Research Management}
As a key measure to promote interdisciplinary cooperation, small research \textbf{teams} are formed to address specific questions. They focus on targeted challenges, such as numerical methods or cooling technologies, and are supported by central funds to organise meetings. A selection of well-defined \textbf{demonstrators}, like the TUG~IM and TUG~PMSM machines of \project{A01}, has been set up to serve as shared experimental benchmarks, enabling projects to validate simulations, refine optimisation workflows and develop advanced cooling techniques. Coherence of joint research across multiple projects is supported by a \textbf{centralised data and software infrastructure}, which enables researchers to work with standardised datasets and computational frameworks, ensuring  reproducibility and minimising redundancy. Reviewed data is publicly shared following the FAIR principles. An \textbf{integrated research training group} organises structured education and training to provide researchers across disciplines with a common framework. It fosters interdisciplinary communication, supports knowledge transfer and promotes scientific exchange through seminars and workshops. Finally, the management board ensures efficient coordination and strategic direction. An advisory board of leading experts provides external guidance, while Mercator Fellows, as renowned international researchers, enhance collaboration by contributing expertise and fostering global networking.

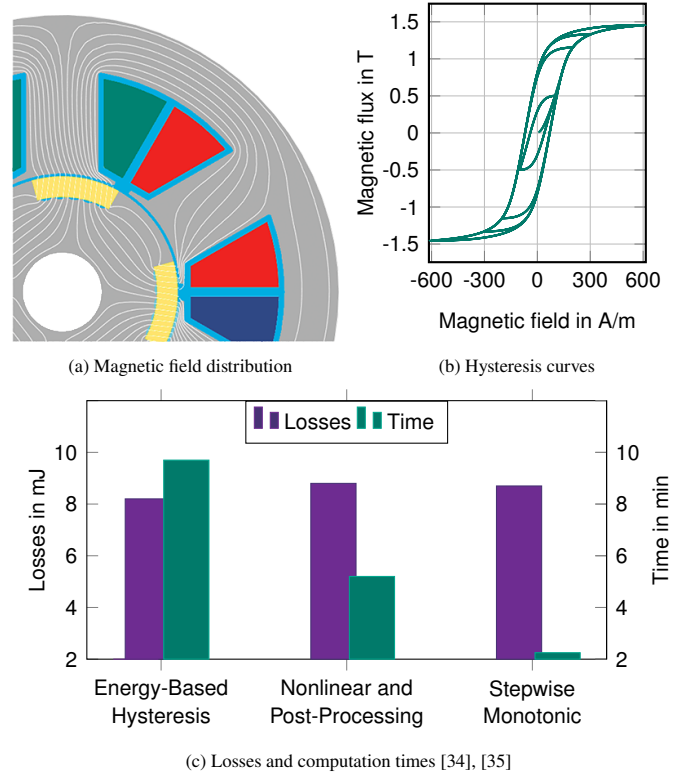
\begin{figure}[t]
    \captionsetup[subfigure]{justification=centering}
    \begin{subfigure}{0.5\linewidth} 
        \includegraphics[height=4.3cm]{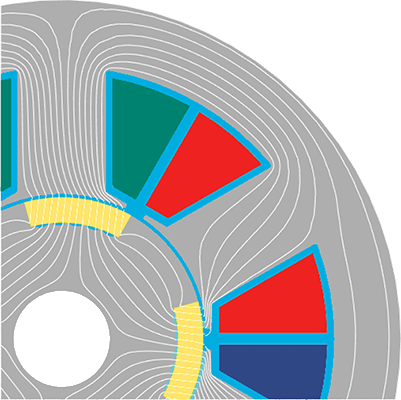}
    \caption{Magnetic field distribution}
    \end{subfigure}%
    \begin{subfigure}{0.5\linewidth}
        \begin{tikzpicture}[font=\sffamily]
            \begin{axis}[
                width=\linewidth,
                height=5cm,
                tick label style={font=\sffamily\footnotesize},
                xlabel style={font=\sffamily\footnotesize},
                ylabel style={font=\sffamily\footnotesize,yshift=-1.2em},
                ylabel={Magnetic flux in T},
                ytick={-1.5,-1,-0.5,0,0.5,1,1.5}, 
                yticklabels={-1.5,-1,-0.5,0,0.5,1,1.5},
                xlabel={Magnetic field in A/m},
                xtick={-600,-300,0,300,600}, 
                xticklabels={-600,-300,0,300,600}, 
                xmin=-615,
                xmax= 615,
                grid=major,
                legend pos=north west,
                legend cell align={left},
                thick,
            ]
            \addplot[color=TUDa-3d,mark=none,smooth] table [col sep=comma,header=false]{figures/success_hysterese1.csv};
            \addplot[color=TUDa-3d,mark=none,smooth] table [col sep=comma,header=false]{figures/success_hysterese2.csv};
            \addplot[color=TUDa-3d,mark=none,smooth] table [col sep=comma,header=false]{figures/success_hysterese3.csv};
            \addplot[color=TUDa-3d,mark=none,smooth] table [col sep=comma,header=false]{figures/success_hysterese4.csv};
            \end{axis}
        \end{tikzpicture}
        \caption{Hysteresis curves}
        \end{subfigure}%
        \\[0.3em]
        \begin{subfigure}{1\linewidth}
    	\centering
        \begin{tikzpicture}
		\begin{axis}[
			ybar,
            width=0.95\linewidth,
            height=5cm,
            tick label style={font=\sffamily\footnotesize},
            xlabel style={font=\sffamily\footnotesize},
            ylabel style={font=\sffamily\footnotesize, anchor=north, at={(axis description cs:0.05,0.5)}},
            yticklabels={0,2,4,6,8,10}, 
            ylabel shift = -5pt,
            bar width=17pt,
			enlarge x limits=0.2,
			bar shift=-5pt,
			ylabel={Losses in mJ},
			symbolic x coords={A, B, C},
            xticklabels={
                {\shortstack{Energy-Based\\Hysteresis}},
                {\shortstack{Nonlinear and\vphantom{g}\\Post-Processing}},
                {\shortstack{Stepwise\\Monotonic}}
            },
            xtick=data,
			xticklabel style={anchor=north},
			ymin=0,
			ymax=10,
			name=plot1
		]
		\addplot+[fill=TUDa-11b,draw=TUDa-11d] coordinates {%
		    (A, 6.2)
		    (B, 6.8)
		    (C, 6.7)
		};
		\end{axis}
		\begin{axis}[
			ybar,
            width=0.95\linewidth,
            height=5cm,
			tick label style={font=\sffamily\footnotesize},
            xlabel style={font=\sffamily\footnotesize},
            ylabel style={font=\sffamily\footnotesize, anchor=north, at={(axis description cs:1.25,0.5)}},
            yticklabels={0,2,4,6,8,10}, 
			bar width=17pt,
			enlarge x limits=0.2,
			symbolic x coords={A, B, C},
            xtick=\empty,
			ymin=0,
			ylabel={Time in min},
			ymax=10,
			axis y line*=right,
			axis x line=none,
            legend style={at={(0.5,1.0)}, anchor=north, legend columns=-1, font=\sffamily\footnotesize},
		]
		\addplot+[fill=TUDa-11d,draw=TUDa-11b] coordinates {(A,0)};
		\addplot+[fill=TUDa-3d,draw=TUDa-3b] coordinates {
			(A, 7.7)
			(B, 3.2)
			(C, 0.25)
		};
		\addlegendentry{Losses}
		\addlegendentry{Time}
		\end{axis}
	   \end{tikzpicture}
    \caption{Losses and computation times \cite{Egger_2025ab,Domenig_2025aa}}
    \end{subfigure}
    \caption{Simulation of hysteresis loops for the TUG PMSM demonstrator.}
    \label{fig:hysteresis}
\end{figure}%

\section{Success Stories}
To support these figures, we highlight selected key research outcomes from the initial funding period, each demonstrating the impact of {interdisciplinary cooperation}.

\subsection{Hysteresis Modelling}
Significant progress was made in the {efficient computation of hysteresis losses in ferromagnetic materials}. 
Projects \nproject{A03}, \nproject{C02} and \nproject{D04} developed a physically consistent and mathematically sound framework for the modelling and simulation of vector hysteresis losses. They studied {calibration to experimental data} and implementation in {advanced finite element solvers} \cite{Sauseng_2024aa,Domenig_2024ac,Egger_2025ac,Egger_2025ab,Domenig_2025aa}.

\begin{figure}
    \centering
    \includegraphics[width=0.9\linewidth]{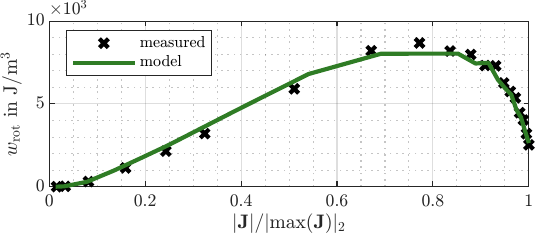}
    \caption{Comparison between the measured and modelled rotational losses.}
    \label{fig:rotLosses}
\end{figure}

Following considerations of \cite{Silvester_1991aa}, the incremental vector hysteresis model of \cite{Francois-Lavet_2013aa} was phrased in the form $\mathbf{b} = \partial_{\mathbf{h}} w_*(\mathbf{h};\{\mathbf{j}_{p,k}\})$, with $\mathbf{b}$ and $\mathbf{h}$ denoting the magnetic induction and field intensity and $\mathbf{j}_{p,k}$ the partial magnetic polarisation of the previous load step. The corresponding {co-energy density}
\begin{align*} 
w_*
= \frac{\mu_0}{2} |\mathbf{h}|^2 - \min\nolimits_{\{\mathbf{j}\}} \sum\nolimits_k \left( u_k(\mathbf{j}_k) - \mathbf{h} \cdot \mathbf{j}_k + \chi_k |\mathbf{j}_k - \mathbf{j}_{p,k}| \right)
\end{align*}
implicitly describes the update $\{\mathbf{j}_{p,k}\} \to \{\mathbf{j}_{k}\}$ of the internal states of the system in response to the magnetic field $\mathbf{h}$, by minimising a certain cost functional depending on the internal energy densities $u_k$ and the pinning strengths $\chi_k$. 
Equivalence with the models of \cite{Francois-Lavet_2013aa,Prigozhin_2016aa} was shown, and a new derivation of the energy-based hysteresis model was obtained, which proves thermodynamic consistency \cite{bookKP_2026}. In doing so, the separation of the field intensity $\mathbf{h}$ into a reversible field $\mathbf{h}_\mathrm{rev}$ and an irreversible field $\mathbf{h}_\mathrm{irr}$ naturally arises from the minimisation principle, meaning these fields do not need to be postulated a priori. In addition, the physically wrong prediction of rotational losses could be resolved by actively rotating the reversible part of the magnetic field strength into the irreversible part. This can be done by determining the angle between those quantities and rotating the reversible part by using a rotational matrix $\mathbf{R}(\varphi)$, where $\varphi$ determines how much rotation is needed and depends on the current polarisation $\varphi = \varphi(\mathbf{j})$ \cite{Domenig_2024aa}. This method is based on determining the reversible and irreversible parts of magnetic field strength from measurements. This provides direct information on how the angle between the two parts behaves, and the parameter occurring in $\varphi(\mathbf{j})$ can be properly fitted to the measurements by the following minimisation
\begin{equation}
    \lvert \mathbf{h}_{\textrm{rev}}(t_k)\rvert = \underset{\lvert \mathbf{h}_{\textrm{rev}}(t_k) \rvert}{\textrm{arg\;min}}\quad \lvert {j}_{\mathrm{an}}(\lvert \mathbf{h}_{\textrm{rev}}(t_k) \rvert) - j^{\mathrm{meas}}(t_k) \rvert_2 \;.
    \label{eq:MKMinHrev}
\end{equation}
In \eqref{eq:MKMinHrev} $j^{\mathrm{meas}}(t_k)$ denotes the magnitude of the measured polarisation for the current time step $t_k$ and $j_{\mathrm{an}}(\lvert \mathbf{h}_{\textrm{rev}} \rvert)$ is the anhysteretic function obtained by uniaxial measurements of the sample. An exemplary result is displayed in \autoref{fig:rotLosses}.
Furthermore, well-posedness and efficient numerical methods could be established, and the corresponding energy functional $w(\mathbf{b};\{\mathbf{j}_{k,p}\}) = \sup_{\mathbf{h}} \mathbf{h} \cdot \mathbf{b} - w_*(\mathbf{h};\{\mathbf{j}_{k,p}\})$ and inverse hysteresis operator $\mathbf{h} = \partial_{\mathbf{b}} w(\mathbf{b};\{\mathbf{j}_{k,p}\})$ could be derived based on arguments of convex analysis. 
A careful calibration of the partial energy densities $u_k$ and pinning forces $\chi_k$ in the model allowed to obtain excellent agreement with measurements. 

Availability of the co-energy density also opened the door to study the well-posedness of the corresponding magnetic field problems, their discretisation by finite element methods and efficient iterative solution by globally convergent generalised Newton-type methods.  
Together with a parallel implementation of the evaluations of $w_*(\mathbf{h};\{\mathbf{j}_{k,p}\})$ and $\partial_{\mathbf{h}} w_*(\mathbf{h};\{\mathbf{j}_{k,p}\})$, we can now simulate every load step of the corresponding magnetic field problem with hysteresis at comparable cost to a standard nonlinear magnetic problem without hysteresis. 
In cooperation with the Mercator Fellows, the methodology was validated for demonstrators `TUDa PMSM' and `TUG PMSM' and compared to different post-processing schemes, leading to the following conclusions:
\begin{itemize}\itemsep-0.2em
\item hysteresis can be included in nonlinear field solvers with minor computational overhead;
\item accurate estimates for hysteresis loss can be computed directly or via post-processing; 
\item further significant speed-up is possible by problem-adapted field extrapolation, see \autoref{fig:hysteresis}.
\end{itemize}
The moderate computational overhead resulting from the efficient realisation of the energy-based vector hysteresis model allows it to be used directly in simulation, as well as for validating various approximations, including standard workflows based on nonlinear field computation and traditional loss formulas.

\subsection{Heatpipes}
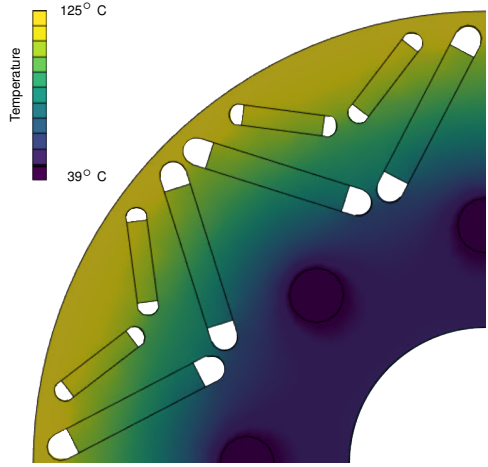
\begin{figure}
    \centering
    \begin{tikzpicture}[font=\tiny\sffamily,inner sep = 0cm]
        \node[anchor=south west, inner sep=0] (image) at (0,0) {%
            \includegraphics[width=6cm]{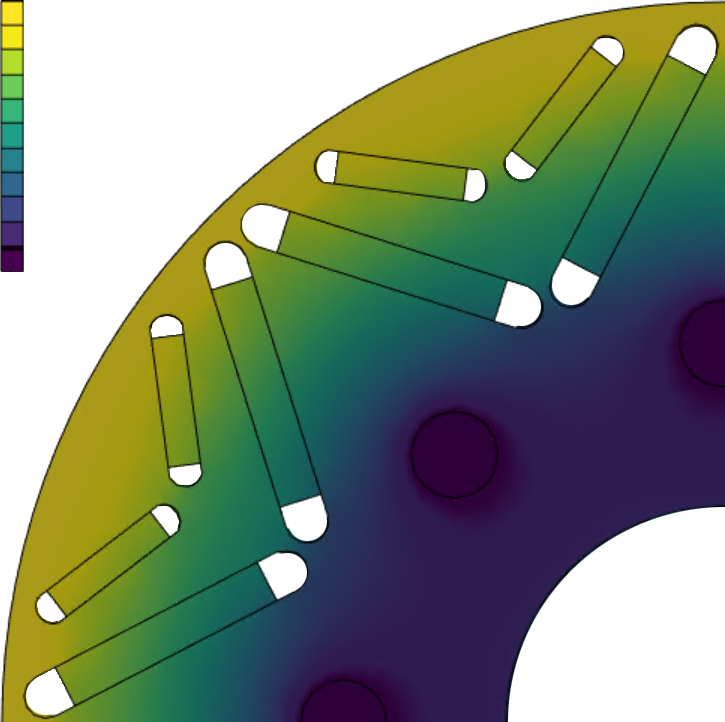}%
        };
        \node[rotate=90] at (-0.2cm,5cm) {Temperature};
        \node[anchor=east] at (1cm,6cm) {125$^\circ$\;C};
        \node[anchor=east] at (1cm,3.8cm) {39$^\circ$\;C};
    \end{tikzpicture}
    \caption{Radial temperature distribution resulting from rotor cooling via multiple off-centre heat pipes.}
    \label{fig:success_heatpipe1}
\end{figure}%

Another key achievement of the first funding period is a novel heatpipe rotor cooling concept developed by Projects \nproject{A06}, \nproject{B02}, \nproject{C02} and \nproject{D02}. The concept is currently evaluated by simulations and will be realised in the demonstrator `TUDa Traction'; a corresponding patent application is pending. The concept remains available to the CRC and will be further investigated by projects and colleagues involved in CSC~4.

PMSMs represent the majority of today’s EV traction motors thanks to their high torque density and drive-cycle efficiency. The rotor-integrated rare-earth magnets must be cooled to avoid exceeding the critical demagnetisation temperature.
Standard cooling solutions for PMSMs are air cooling (railway, wind power, industrial motors), stator water jacket and rotor water lance cooling in a rotor shaft (e.g.\ Mercedes EQS or EQC) and oil flow through the rotor (e.g.\ VW APP550, Tesla Model Y) or spraying at the axial rotor ends, often as a side-effect of stator winding oil spray cooling \cite{Gronwald_2021aa}. Air cooling is undesirable in the automotive sector due to sealing requirements. 

\begin{figure}
    \captionsetup[subfigure]{justification=centering}
    \centering
    \begin{subfigure}[c]{.25\linewidth}
        \centering
        \includegraphics[width=\linewidth]{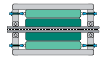}
        \caption{Stator spray\\cooling.}
        \label{fig:ssc}
    \end{subfigure}%
    \begin{subfigure}[c]{.25\linewidth}
        \centering
        \includegraphics[width=\linewidth]{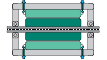}
        \caption{Stator jet\\impingement cooling.}
        \label{fig:sjic}
    \end{subfigure}%
    \begin{subfigure}[c]{.25\linewidth}
      \centering
      \includegraphics[width=\linewidth]{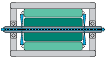}
      \caption{Radial rotor spray cooling.}
      \label{fig:rrsc}
    \end{subfigure}%
    \begin{subfigure}[c]{.25\linewidth}
      \centering
      \includegraphics[width=\linewidth]{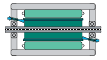}
      \caption{Rotor heatpipe with spray cooling.}
      \label{fig:rjic}
    \end{subfigure}\\
    \caption{Cooling concepts visualised following Gronwald et al. \cite{Gronwald_2021aa} (with permission from IEEE).}
    \label{fig:cooling_types}
\end{figure}

With water-lance cooling, the radial heat path from the magnets to the coolant in the shaft is long, making cooling less effective. Rotating seals to the engine interior are also required. With direct oil flow, the rotor concept becomes very complex due to the oil channels in the laminated core, with high demands on package tightness etc. 
With today's common oil-spray cooling of the stator winding, either from housing parts or out of holes in the shaft, only the axial rotor ends are in contact with oil. This results in higher temperatures in the axial centre of the stack. The target is to reduce the thermal resistance from the magnets to the coolant. A reduced thermal resistance results in a lower temperature difference between coolant and magnets. In the rotor stack, heat conduction is the dominant effect, where the thermal resistance can be expressed as
\begin{equation}
    R_{\mathrm{th}}=\frac{l_{\mathrm{th}}}{\lambda_{\mathrm{core}} A}
\end{equation}
with the thermal conductivity $\lambda_{\mathrm{core}}$ of the rotor lamination in W/(m K), the length of the thermal path $l_{\mathrm{th}}$ and the cross section of the path $A$. With a shaft cooling, the cross section narrows from the magnet position to the shaft, leading to a high thermal resistance combined with a high temperature drop in the region with lower radius, see~\autoref{fig:success_heatpipe1}.
Using an oil spray against the axial ends, the thermal path is in axial direction. Due to the stack of thin, insulated iron sheets in the axial direction, the thermal conductivity is typically one order of magnitude lower than in the radial direction, leading to a high thermal resistance as well.
To overcome these drawbacks, we developed a {concept with rotor-integrated heat pipes} in the axial direction, realising a strongly reduced thermal resistance. %
A careful arrangement of the heat pipes in the vicinity of the permanent magnets enables highly efficient heat transport to the axial rotor ends. By adding an {oil spray to the axial rotor ends}, the heat pipes can be back-cooled, dissipating heat from the rotor core; see \autoref{fig:rjic}. To maximise the heat transfer coefficient from the heat pipes to the oil, aluminium rotor end plates are foreseen, representing a larger spray target for the oil and distributing the heat to be dissipated over a larger surface.
In conclusion, a {highly effective rotor cooling solution} was established which is easy to manufacture.

\subsection{Isogeometric Analysis}
\begin{figure*}
    \centering
    \captionsetup[subfigure]{justification=centering}
    \begin{subfigure}[t]{.22\linewidth}
      \centering
    \includegraphics[width=\linewidth]{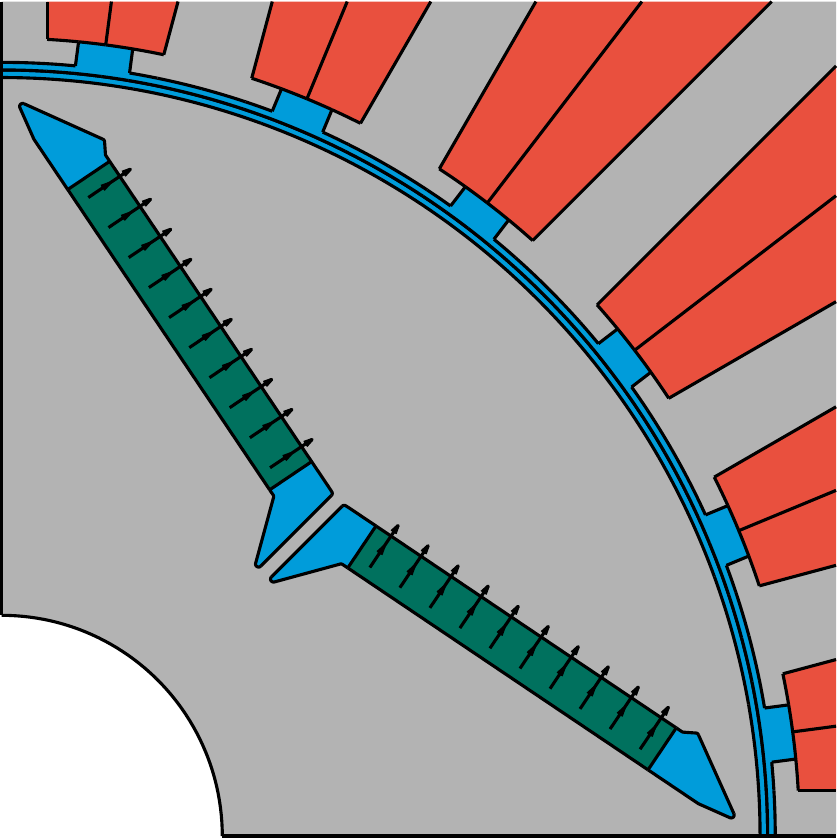}
        \caption{Parameter optimised.}
      \label{fig:Results:ParamOpt}
    \end{subfigure} \hfill
    \begin{subfigure}[t]{.22\linewidth}
      \centering
        \includegraphics[width=\linewidth]{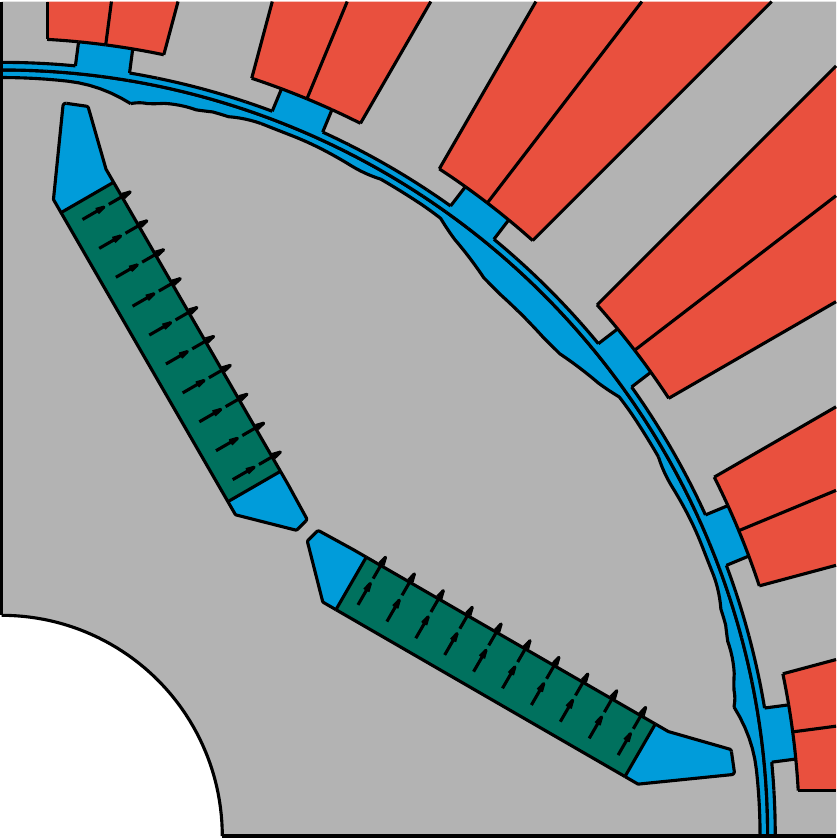}
      \caption{Shape optimised.}
      \label{fig:Results:ShapeOpt}
    \end{subfigure} \hfill
    \begin{subfigure}[t]{.22\linewidth}
      \centering
      \includegraphics[width=\linewidth]{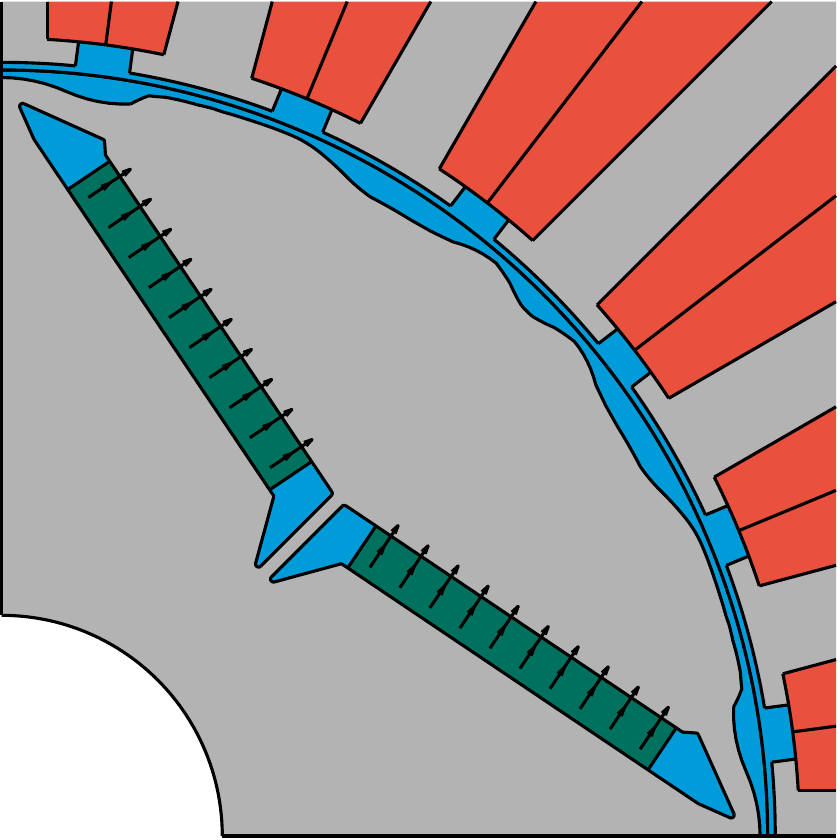}
      \caption{Parameter then shape.}
      \label{fig:Results:ParamThenShapeOpt}
    \end{subfigure} \hfill
    \begin{subfigure}[t]{.22\linewidth}
      \centering
      \includegraphics[width=\linewidth]{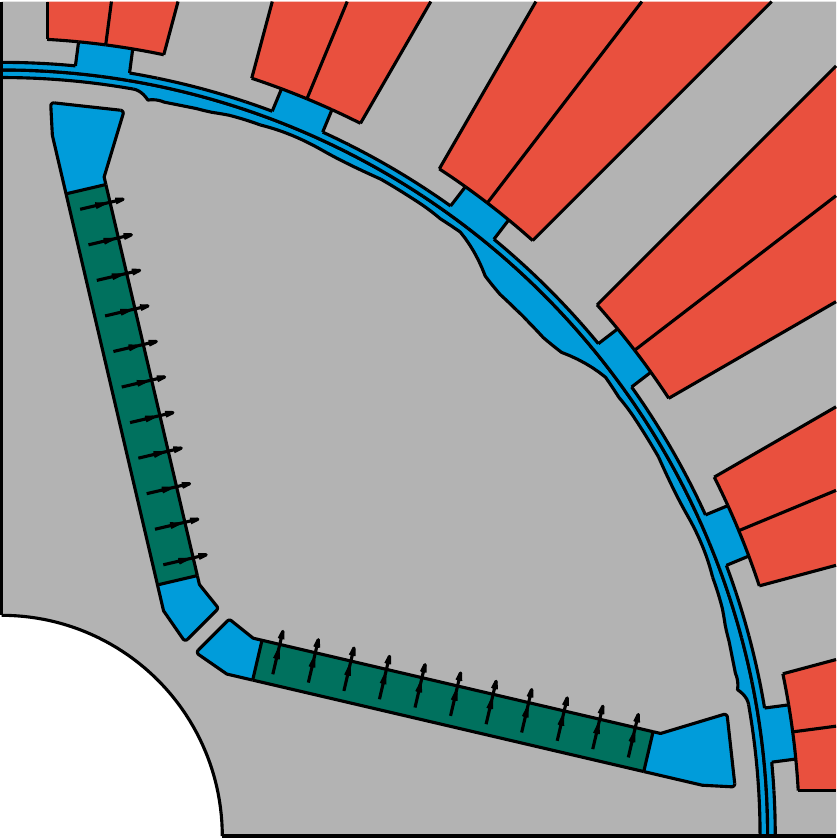}
      \caption{Parameter and shape.}
      \label{fig:Results:ParamShapeOpt}
    \end{subfigure}
    \caption{Optimisation results of different approaches for the JMAG PMSM demonstrator, see \cite{Wiesheu_2024aa}.}
    \label{fig:Results:OptComparison}
\end{figure*}
Isogeometric Analysis (IGA) can be seen as an extension of the conventional Finite Element technique by B-Splines and Non-Uniform Rational B-Splines (NURBS). B-Splines and NURBS are the standard modelling tools when it comes to creating geometries with modern Computer-Aided Design (CAD) software. Using B-Splines and NURBS for both the geometry description and as basis for the simulation with IGA holds the promise of solving many issues that machine designers face today. For example, parameter optimisation workflows typically rely on genetic algorithms and repetitive meshing. Shape optimisation, on the other hand, uses gradients, but the optimised vertices must be reverse engineered in the CAD program. 

Against this background, a comprehensive {spline-based simulation and optimisation workflow} 
for electric machine models has been jointly developed in the project. 
It combines {several key advances} in spline-based geometry handling, isogeometric and harmonic mortaring, shape and parameter derivatives and gradient-based optimisation \cite{Merkel_2022ab,Ziegler_2023ab,Komann_2024aa,Wiesheu_2024aa}. 

State-of-the-art industrial workflows for optimisation start from a machine template, e.g.\ a single-V topology with a particular winding scheme for the stator, and then perform parameter-based optimisation of width, length and radii. For each parameter combination, a new model, typically based on lines and arcs, is constructed, meshed and then solved using the finite element method. Finally, the relevant Quantities of Interest (QoIs) are returned to the optimiser. 

In contrast, we construct geometries using {volumetric representations} where the spline mapping to the physical space is prescribed by control points. By using splines, variations of the control points are automatically propagated to the underlying finite-element description, avoiding the need for remeshing or mesh morphing.
If the same high-order splines are also used for the solution representation, the number of degrees of freedom can also be kept small due to the excellent approximation properties of the basis functions. In addition to the geometry parameters, which define the initial position of the control points according to the geometry template, the control points may be directly adjusted. 

Due to the continuous mapping, gradients with respect to the control points and associated parameters are straightforward to compute and computationally efficient to obtain. This allows for optimisation routines with many design parameters and shape-defining control points, {unlocking designs beyond the original template}. 

The method has been successfully demonstrated on a permanent magnet synchronous motor model from JMAG, achieving {significant reductions in magnet mass and torque ripple} while maintaining performance constraints. Exemplary results are seen in  \autoref{fig:Results:OptComparison}, where the optimisation is compared for sequential and combined parameter and shape optimisation.

Compared to conventional optimisation techniques, the combination of parameter and free-form optimisation results in:
\begin{itemize}\itemsep-0.2em
    \item significant reduction in magnet material usage, reducing costs and environmental impact;
    \item almost complete elimination of torque ripples, improving efficiency and reducing vibration;
    \item computationally efficient solvers that are competitive with other open-source and commercial tools;
    \item an overall accelerated workflow with fast gradient evaluations and no remeshing overhead;
    \item directly usable CAD geometry, which does not need to be reconstructed;
\end{itemize}
These results demonstrate that IGA is numerically advantageous and has clear potential for industrial applications.

\subsection{Topology Optimisation}
In a cooperation among projects \nproject{A01}, \nproject{A02}, \nproject{A03}, \nproject{C02} and \nproject{D02}, a {multi-material design optimisation} framework for the {integrated multiphysical} topology optimisation of two-dimensional models of electric machines has been developed. This research endeavour requires accurate mathematical modelling, efficient simulation techniques and a flexible optimisation method.

The conventional machine design workflow in industry accounting for multiple physical domains is a {sequential} one where, e.g. thermal networks are extracted based on the original design, but not updated during electromagnetic design optimisation, and thermal optimisation is carried out afterwards. Here, we contribute to a {shift in paradigm} by performing {integrated electromagnetic-thermal-mechanical} topology optimisation while also {avoiding demagnetisation} of the permanent magnets. 

\begin{figure*}
	\captionsetup[subfigure]{justification=centering}
	\centering
	\begin{subfigure}[c]{.16\linewidth}
		\centering
		\includegraphics[width=.95\linewidth, trim=850 40 700 425, clip]{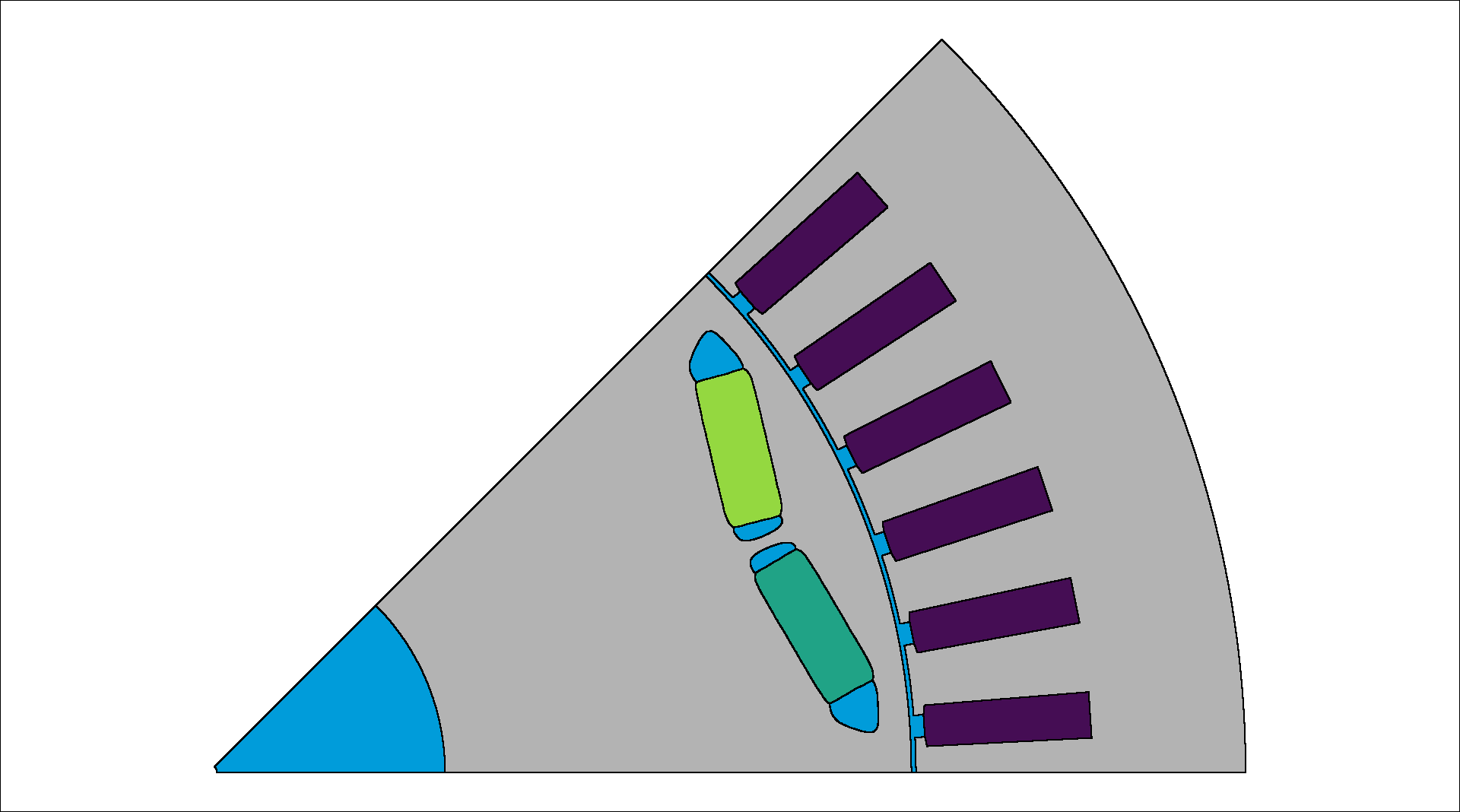}
        \caption{Reference design}
	\end{subfigure}%
    \hfill
	\begin{subfigure}[c]{.16\linewidth}
		\centering
		\includegraphics[width=.95\linewidth, trim=850 40 700 425, clip]{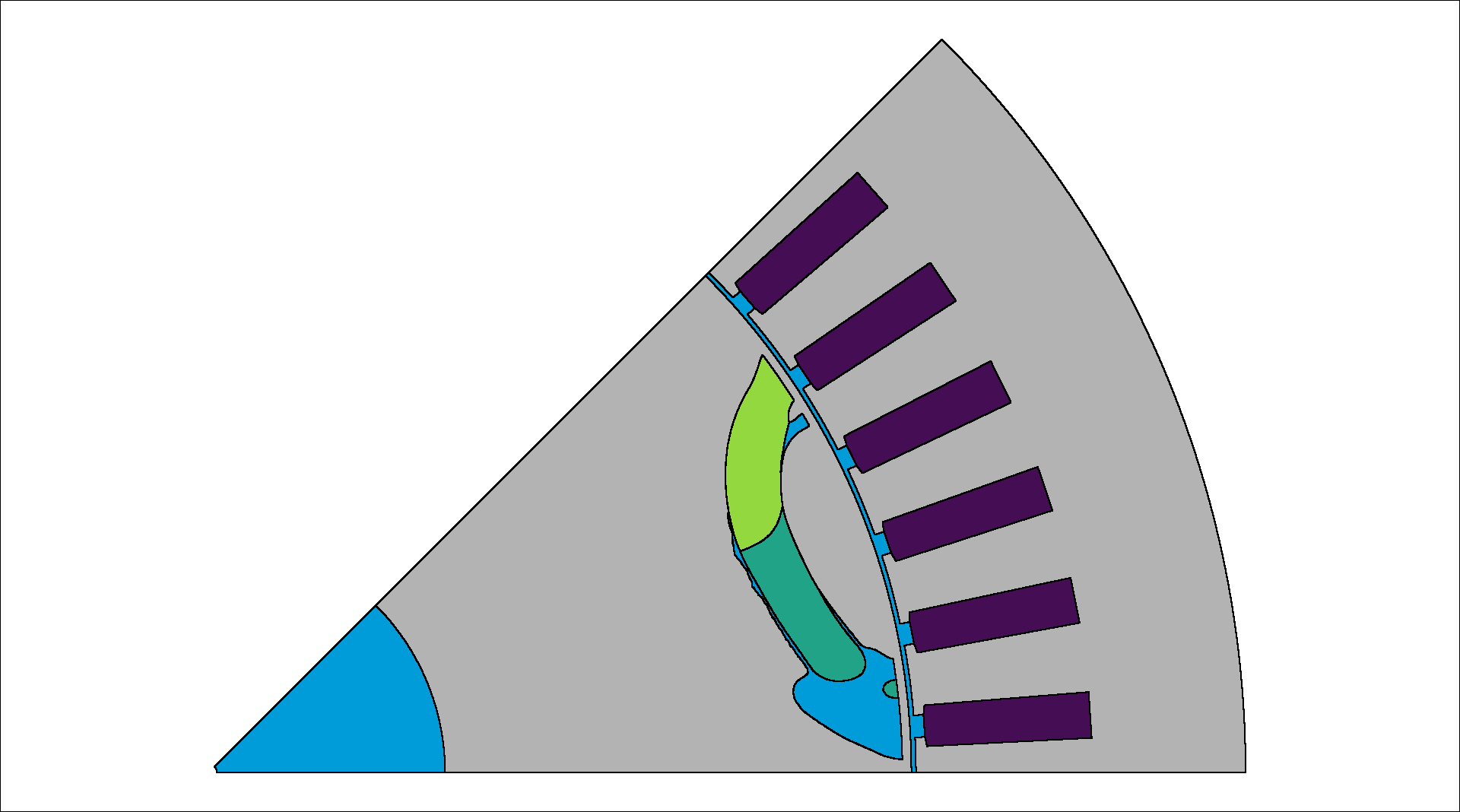}
        \caption{Unconstrained}
	\end{subfigure}%
    \hfill
	\begin{subfigure}[c]{.16\linewidth}
		\centering
		\includegraphics[width=.95\linewidth, trim=850 40 700 425, clip]{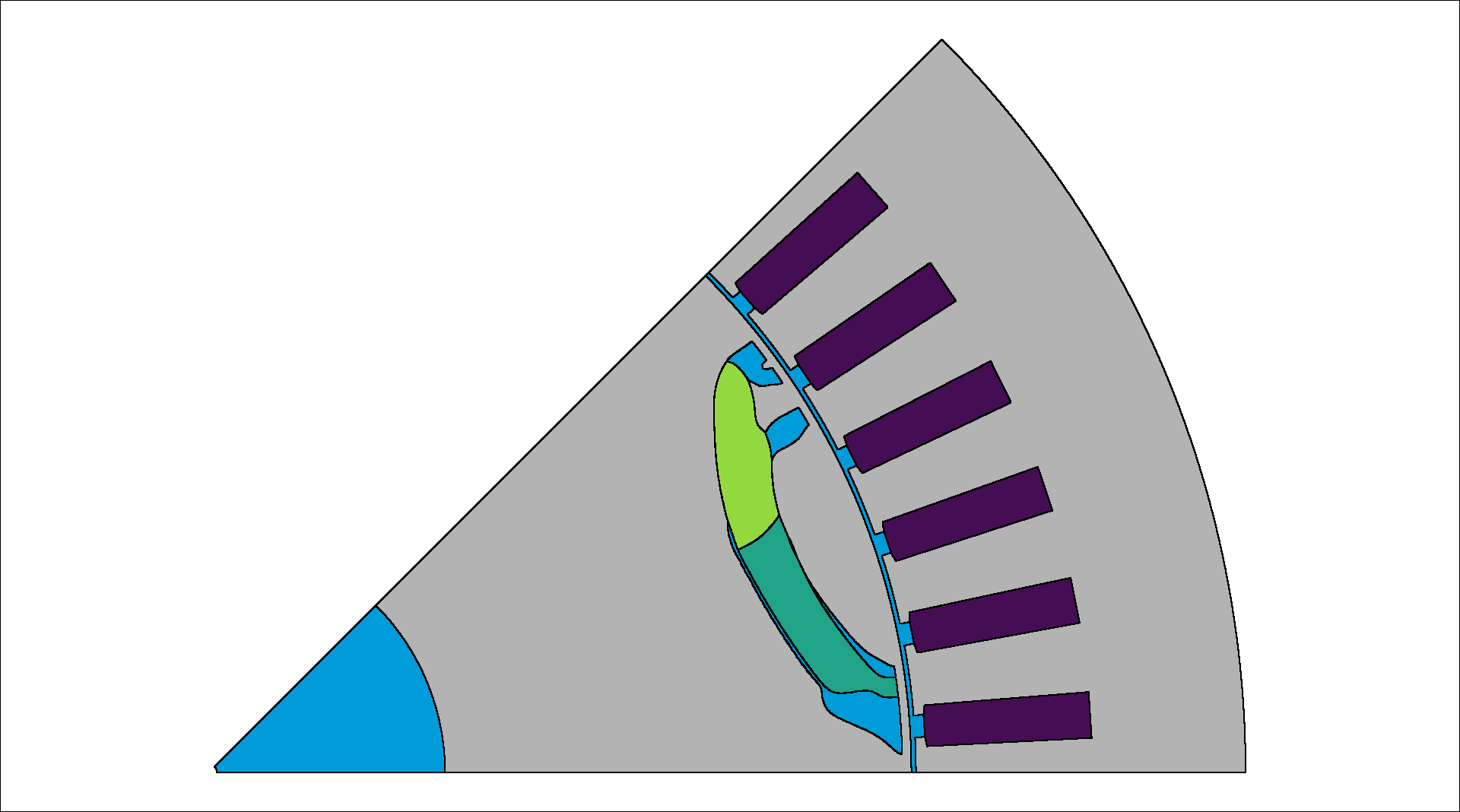}
        \caption{Temperature constraint}
	\end{subfigure}%
    \hfill
	\begin{subfigure}[c]{.16\linewidth}
		\centering
		\includegraphics[width=.95\linewidth, trim=850 40 700 425, clip]{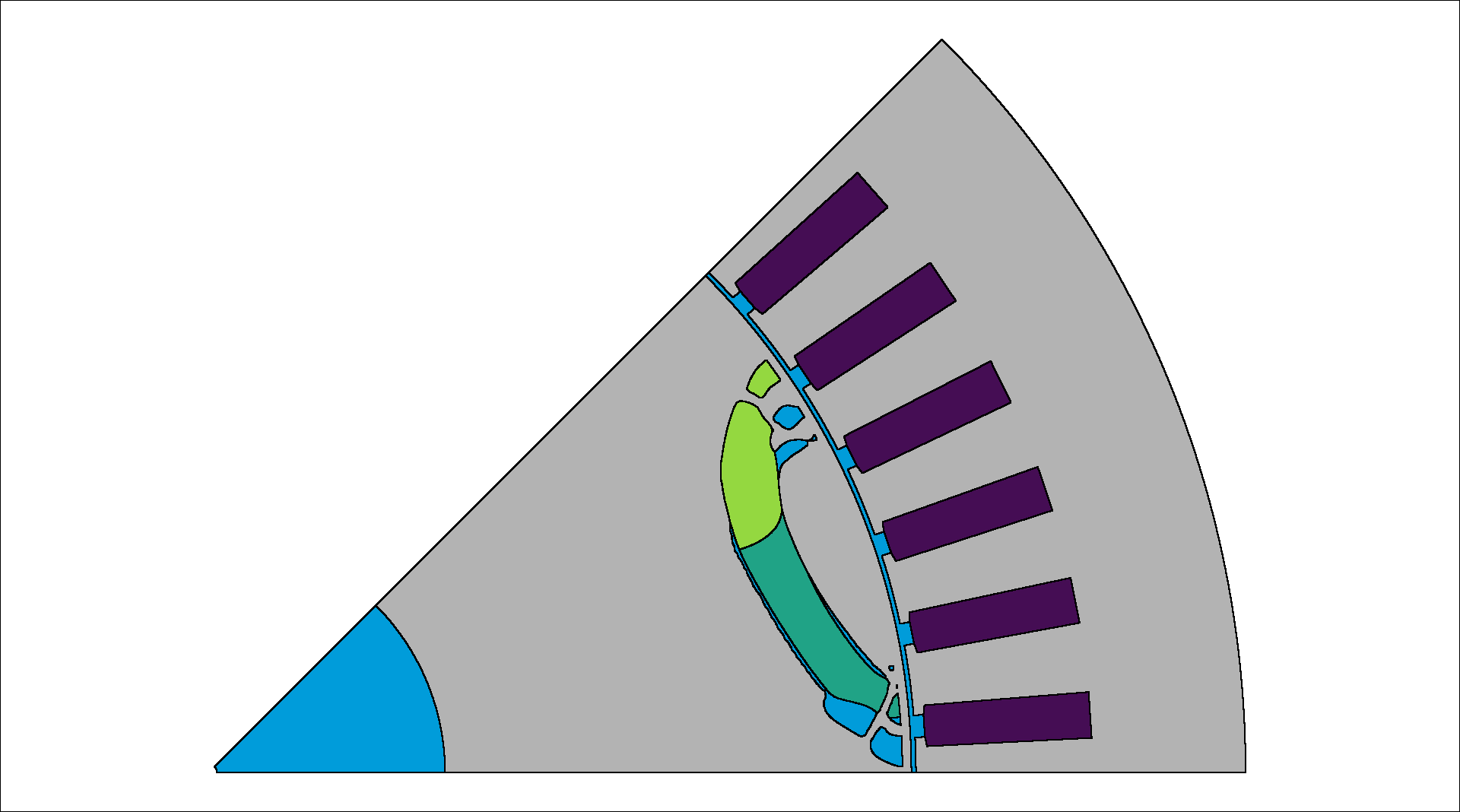}
        \caption{Stress constraint}
	\end{subfigure}%
    \hfill
	\begin{subfigure}[c]{.16\linewidth}
		\centering
		\includegraphics[width=.95\linewidth, trim=850 40 700 425, clip]{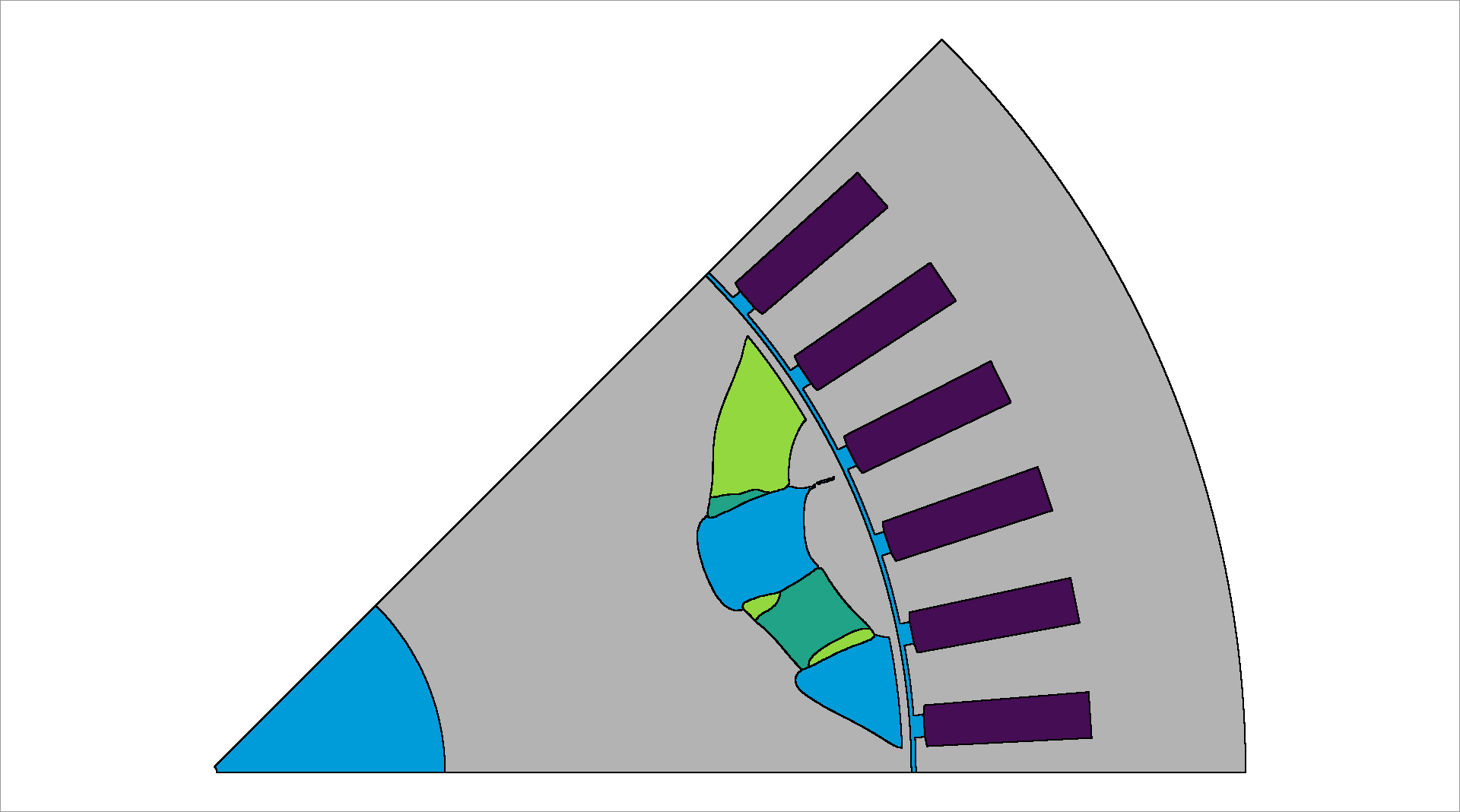}
        \caption{Demagn. constraint}
	\end{subfigure}%
    \hfill
	\begin{subfigure}[c]{.16\linewidth}
		\centering
		\includegraphics[width=.95\linewidth, trim=850 40 700 425, clip]{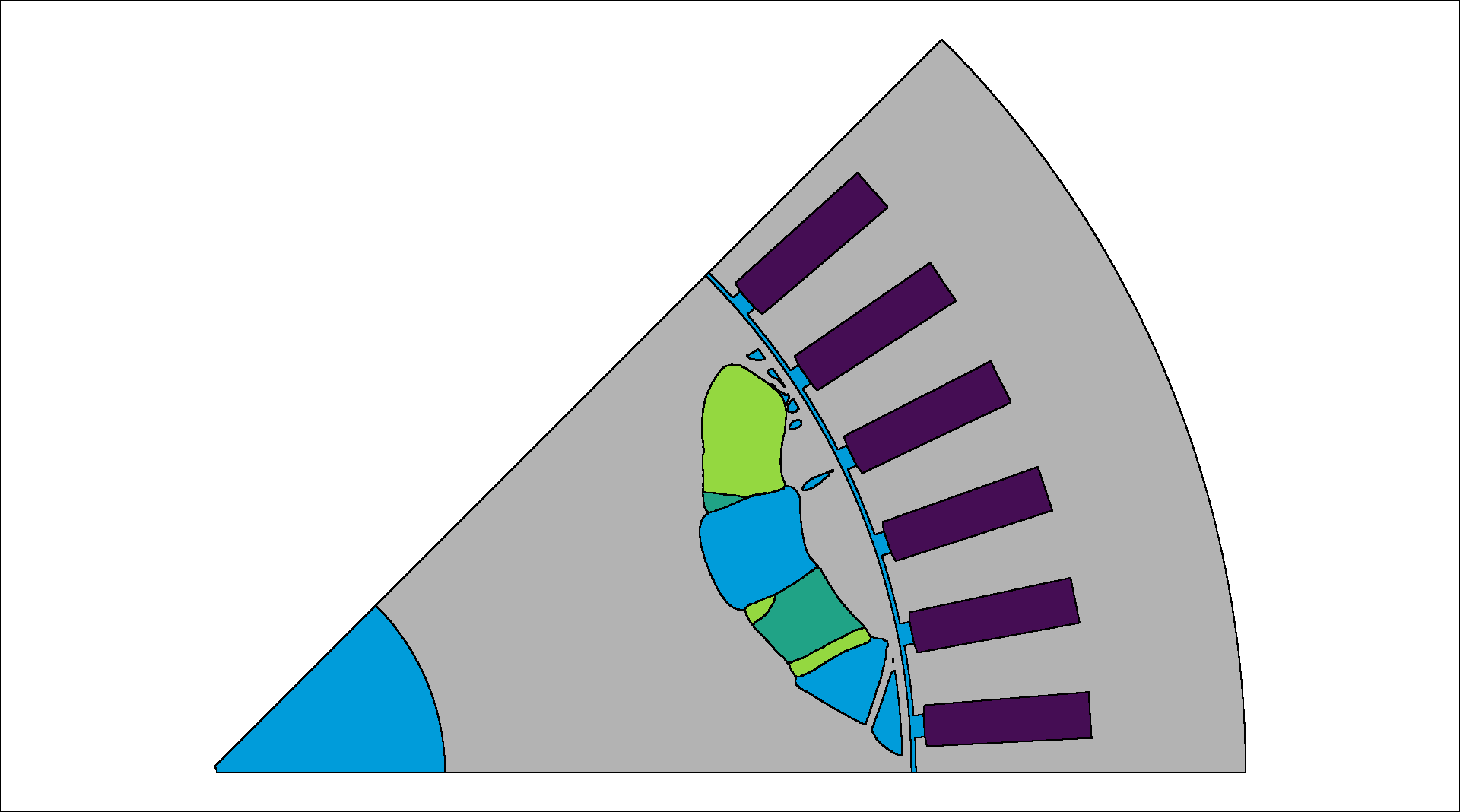}
        \caption{All constraints}
	\end{subfigure}
	\caption{Conventional machine design (a), electromagnetic torque $\mathcal{T}$ optimisation without (b) and with additional constraints: temperature constraint $\mathcal{C}_\theta$ (c), stress constraint $\mathcal{C}_\sigma$ (d), demagnetisation constraint $\mathcal{C}_d$ (e) and temperature, stress and demagnetisation constraint (f).}
    \label{fig_highlightTopOpt}
\end{figure*}

Besides the electromagnetic equations, this problem also involves a heat equation accounting for eddy current losses in permanent magnets and a linear elasticity problem accounting for centrifugal forces. The time-periodic nonlinear magneto-quasi-static problem accounting for a nonlinear material law in the permanent magnets was efficiently solved by the harmonic mortar method \cite{Egger_2022ab}. Finally, the optimisation was based on the concept of {topological derivatives} \cite{Gangl_2022aa} and a {vector-valued level set} framework \cite{Gangl_2020ac}, allowing for the optimal placement of an arbitrary number of materials (e.g.\ iron, air, magnets with different magnetisation directions) having well-defined, crisp material interfaces.

As an illustrative example, we considered a permanent magnet synchronous machine as depicted in \autoref{fig_highlightTopOpt}(a). We maximised the {average torque} $\mathcal T$ by distributing iron, air and magnets with two different magnetisation directions in the rotor while keeping the total magnet volume fixed. Also, we considered
\begin{itemize}\itemsep-0.2em
    \item a local {temperature} constraint $\mathcal{C}_\theta := \theta(x) / \theta^* \leq 1$ in the permanent magnets with the maximum allowed temperature $\theta^* = 90^\circ$C,
    \item a local {stress} constraint $\mathcal{C}_\sigma := \sigma_{\mathrm{VM}}(x) / \sigma^* \leq 1$ in the rotor with the maximum allowed von Mises stress $\sigma^* = 500$MPa,
    \item and a local {demagnetisation} constraint $\mathcal{C}_d := (B^0-B(x) \cdot e_\varphi) / (B^0 - B^*) \leq 1$ implying a lower bound on the flux density in the magnetisation direction, $B(x) \cdot e_\varphi \geq B^* = 0.56$T with the remanence flux density $B^0=1.216$T used for scaling.
\end{itemize}

\begin{figure}
    \vspace*{-0.3cm}
    \centering
    \includegraphics[width=.45\textwidth]{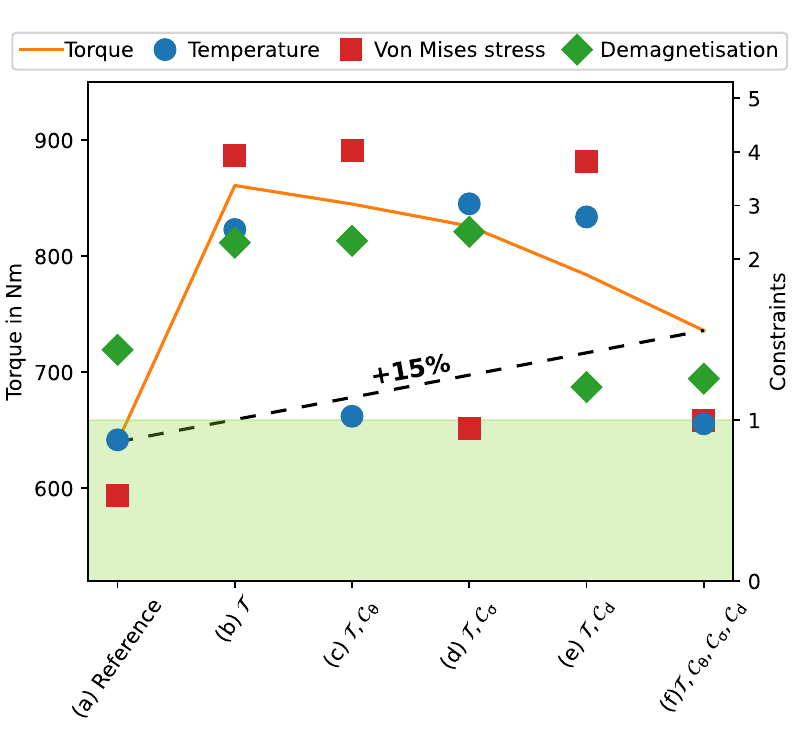}\\[-0.6cm]
    \caption{Evaluation of the designs in \autoref{fig_highlightTopOpt} in terms of torque and constraints on maximum temperature, von Mises stress and demagnetisation. Constraints are satisfied in the highlighted area.}
    \label{fig_highlightTop_plot}
    \vspace*{-0.3cm}
\end{figure}

\autoref{fig_highlightTopOpt} depicts the conventional machine design and optimisation results without additional constraints, with each of the three constraints considered separately, and with all constraints combined. Figure \ref{fig_highlightTop_plot} shows the torque value for all these designs along with the evaluation of the local constraints. 
The unconstrained optimisation result of \autoref{fig_highlightTopOpt}(b) yields an increase in torque of around 35\%, however all constraints are violated significantly. Imposing a local temperature constraint leads to a design where the magnets are further away from the air gap with a maximum temperature of 92$^\circ$C, see \autoref{fig_highlightTopOpt}(c). The stress constraint leads to iron bridges connecting the outer iron in \autoref{fig_highlightTopOpt}(d) and the demagnetisation constraint leads to divided and thicker magnets, see \autoref{fig_highlightTopOpt}(e). By imposing all constraints, we obtain the design of \autoref{fig_highlightTopOpt}(f) where all the described effects are combined. The design approximately satisfies all constraints while giving a 15\% increase in torque. The fact that some constraints are only approximately satisfied is due to the chosen penalty formulation with large, but finite weights.

The problem was jointly defined with Robert Bosch GmbH. The results in \autoref{fig_highlightTopOpt} consider only a single operating point, which results in asymmetric designs. 
The extension of the presented approach from considering only a fixed operating point to optimising its performance over a full drive cycle is the subject of current work, with first results available in \cite{Krenn_2025ab}. 

\section{Conclusion}
The first funding period of the CREATOR project has demonstrated the feasibility and potential of a fully integrated, simulation-driven approach to electric machine design. By bringing together expertise from electrical engineering, applied mathematics, fluid dynamics and materials science, the project has established novel methodologies that address long-standing challenges in the field.

The establishment of shared demonstrators, standardised datasets and a centralised software infrastructure has fostered interdisciplinary collaboration across the project. These foundations ensure reproducibility, minimise redundancy and accelerate scientific progress throughout the CRC.

Looking ahead to the second funding period, CREATOR is well positioned to tackle emerging challenges. The expanded focus on three-dimensional effects, particularly in axial flux machines, the increased emphasis on machine learning and data-driven techniques, and the introduction of advanced cooling strategies all reflect the evolving demands of high-performance electric machine design. The inclusion of new projects and Mercator Fellows strengthens the CRC's capacity to address these complex, multiphysical problems.

We hope that this article will serve as inspiration and provide useful guidance for the development of similar projects in the community. We would greatly appreciate comments and feedback from colleagues.

\printbibliography


\section*{Author Names and Affiliations}

\textbf{Sebastian Schöps.} Computational Electromagnetics Group. Department of Electrical Engineering and Information Technology, Technische Universität Darmstadt, Darmstadt, 64289 Darmstadt, Germany. \href{mailto:sebastian.schoeps@tu-darmstadt.de}{sebastian.schoeps@tu-darmstadt.de}.

\textbf{Annette Mütze.} Institute of Electrical Power Systems. Graz University of Technology, 8010 Graz, Austria. \href{mailto:annette.muetze@tugraz.at}{annette.muetze@tugraz.at}.

\textbf{Herbert De Gersem.} Institute for Accelerator Science and Electromagnetic Fields. Department of Electrical Engineering and Information Technology, Technische Universität Darmstadt, Darmstadt, 64289 Darmstadt, Germany. \href{mailto:herbert.degersem@tu-darmstadt.de}{herbert.degersem@tu-darmstadt.de}.

\textbf{Herbert Egger.} Institute for Analysis and Scientific Computing, Johannes Kepler University Linz, 4040 Linz, Austria, and Johann Radon Institute for Computational and Applied Mathematics (RICAM), Austrian Academy of Sciences, 4040 Linz, Austria. \href{mailto:herbert.egger@jku.at}{herbert.egger@jku.at}.

\textbf{Manfred Kaltenbacher.} Institute of Fundamentals and Theory in Electrical Engineering. Graz University of Technology, 8010 Graz, Austria. \href{mailto:manfred.kaltenbacher@tugraz.at}{manfred.kaltenbacher@tugraz.at}.

\textbf{Markus Lazanowski.} Profile topic Computational Engineering, Energy and Environment (E+E), Technische Universität Darmstadt, Darmstadt, 64289 Darmstadt, Germany. \href{mailto:markus.lazanowski@tu-darmstadt.de}{markus.lazanowski@tu-darmstadt.de}.

\bigskip

\textbf{CREATOR Team}: the following colleagues are members of Technische Universität Darmstadt (Germany):
Yves Burkhardt,
Andreas Dreizler,
Jeanette Hussong,
Florian Kummer,
Melina Merkel,
Martin Oberlack,
Ilia Roisman,
Yvonne Späck-Leigsnering,
Stefan Ulbrich,
Oliver Weeger,
Bai-Xiang Xu,
of Graz University of Technology (Austria):
Günter Brenn,
Benjamin Marussig,
Carole Planchette,
Martin Schanz,
Olaf Steinbach,
and of RICAM (Austria):
%
Peter Gangl.

\end{document}